\documentclass[a4paper,11pt]{article}
\pdfoutput=1
\usepackage{jheppub}

\usepackage{slashed}
\usepackage{bm,wasysym}
\usepackage[normalem]{ulem}
\usepackage{amsmath,amsfonts,slashed,amssymb,tikz,bm,psfrag,graphicx,color,dsfont}

\def\slash#1{\setbox0=\hbox{$#1$}\dimen0=\wd0
      \setbox1=\hbox{/} \dimen1=\wd1 \ifdim\dimen0>\dimen1
      \rlap{\hbox to \dimen0{\hfil/\hfil}} #1                        \else
      \rlap{\hbox to \dimen1{\hfil$#1$\hfil}}
      /   \fi}

\newcommand{\lsim}{
\mathrel{\hbox{\rlap{\hbox{\lower4pt\hbox{$\sim$}}}\hbox{$<$}}}}

\newcommand{\gsim}{
\mathrel{\hbox{\rlap{\hbox{\lower4pt\hbox{$\sim$}}}\hbox{$>$}}}}

\allowdisplaybreaks[2]

\newcommand{\SCETI}{SCET$_\text{I}$}
\newcommand{ \LambdaQCD}{ \Lambda_{\mathrm{QCD}}}

\title{Complete analysis on QED corrections to $B_{q} \, \to\, \tau^+\,  \tau^-$}

\author[a]{Yong-Kang Huang,}
\author[b]{Yue-Long Shen,}
\author[a]{Xue-Chen Zhao,}
\author[c*]{and Si-Hong Zhou}

\affiliation[a]{School of Physics, Nankai University, 300071 Tianjin, P.R. China }
\affiliation[b]{ College of Information Science and Engineering,
Ocean University of China,\newline
Qingdao, 266100 Shandong, P.R. China}
\affiliation[c]{ School of Physical Science and Technology,
Inner Mongolia University,,\newline
Hohhot 010021, P.R. China }

\emailAdd{2120190131@mail.nankai.edu.cn}
\emailAdd{shenylmeteor@ouc.edu.cn}
\emailAdd{zxc@mail.nankai.edu.cn}
\emailAdd{shzhou@imu.edu.cn}
\abstract{
Motivated by a dynamical enhancement of the electromagnetic corrections 
by a power of $\Lambda_{\mathrm{QCD}}/m_b$ in $B_{d,s}\, \to\, \mu^+\, \mu^-$ 
at next-to-leading order (NLO), we extend the QED factorization effects on the 
leptonic $B$ meson decays with light muon leptons to tauonic final states, 
$B_{d,s} \, \to\, \tau^+\, \tau^-$, using soft-collinear effective theory (SCET). 
This extension is necessary owing to the appearance of the large $\tau$ mass, 
which will lead to different power counting in SCET and also different results. 
We provide a complete NLO electromagnetic corrections to 
$B_{d,s} \, \to\,  \tau^+\, \tau^-$, which include hard functions and hard-collinear 
functions below the bottom quark mass scale $\mu_b$. The power enhanced 
electromagnetic effects from hard-collinear contributions on 
$B_{d,s}\, \to\,  \mu^+\,  \mu^-$ discussed before also exist in 
$B_{d,s} \, \to\,  \tau^+\,  \tau^-$. However the logarithm term arising from 
contributions of hard-collinear photon and lepton virtualities for 
$B_{d,s}\,  \to\,  \tau^+\,  \tau^-$ is not large as it is in muon case due to 
the hard-collinear scale of $\tau$ mass, which lead to only approximately 
$0.04\%$ QED corrections to the branching fraction of 
$B_{d,s} \, \to\,  \tau^+\,  \tau^-$ compared with overall reduction about 
$0.5\%$ in $B_{d,s}\, \to\,  \mu^+\,  \mu^-$. } 

\note{*corresponding author: shzhou@imu.edu.cn}

\begin{document}

\maketitle


\section{Introduction}
\label{Introduction}
 
The purely leptonic decays $B_q \, \to\,  \ell^+\,  \ell^-$, with $q=d,\, s$ 
and $\ell= e, \mu, \tau$, are highly suppressed in the standard model (SM) 
due to the loop suppresses (FCNC) and helicity suppresses. Therefore, 
they have an important role in the study of physics beyond the Standard 
Model (BSM). They are also of interest owning to their clear theoretical 
descriptions. In fact, the only relevant quantity that needs to be calculated 
at the leading order of $\alpha_{\mathrm{em}}$ is $B$-meson decay constant 
$f_B$. The branching ratio of $B_q\,  \to\,  \ell^+ \, \ell^-$ at the leading orders 
in flavor-changing weak interactions and in $m_{B_{q}}^{2} / m_{W}^{2}$ can 
be expressed as \cite{Bobeth:2013uxa},
   \begin{eqnarray} \label{alphaem0}
   {\mathcal{B}}\left[\, B_{q} \, \rightarrow \, \ell^{+} \, \ell^{-}\, \right]\, =\, 
   \frac{|N|^{2} m_{B_{q}}^{3}\,  f_{B_{q}}^{2}}{8\,  \pi\,  \Gamma_{H}^{q}}\,
   \beta_{q \ell}\, r_{q \ell}^{2}\, \left|C_{A}\left(\mu_{b}\right)\right|^{2}
   \, +\, \mathcal{O}\left(\alpha_{\mathrm{em}}\right)\, ,
   \end{eqnarray}
where $r_{q \ell}=2 m_{\ell} / m_{B_{q}}$ and 
$\beta_{q \ell}=\sqrt{1- r_{q \ell}^{2}}$. The normalization constant 
$N =V_{t b}^{*} V_{t q} G_{F}^{2} m_{W}^{2} / \pi^{2}$ and 
$\Gamma_{H}^{q}$ denotes the heavier mass-eigenstate total width 
of $B_{q}-\bar{B}_{q}$ mixing. $C_{A}$ is the 
$\overline{\mathrm{MS}}$-renormalized Wilson coefficient associated with 
the operator $\left[\, \bar{b}\,  \gamma_{\alpha}\,  \gamma_{5} \, q\, \right]\, 
\left[\, \bar{\ell} \, \gamma^{\alpha}\,  \gamma_{5}\,  \ell\, \right]$ at the scale 
$\mu_{b}$. Up to date, the most precise determinations of $f_{B_{u,d}}$ 
and $f_{B_s}$ have already reached the relative precision of about $0.7\%$ 
and $0.5\%$ from lattice QCD, respectively \cite{Bazavov:2017lyh}. 
These precise values of $f_B$ from lattice calculations provide a motivation 
for improving the perturbative ingredients which arise from several energy 
scales spanned by the SM. The QCD corrections to $C_{A}$ have been up 
to the next-to-next-to-leading order (NNLO) \cite{Hermann:2013kca}. At the 
scale $\mu\,  \geq \, \mu_b$, the electroweak (EW) corrections at NLO have 
been done in \cite{Bobeth:2013tba}, which combined with NNLO QCD 
corrections are calculated in \cite{Bobeth:2013uxa}. In recent years, a 
consistent simultaneous treatment of QCD and QED corrections to 
$B_{q}\, \to\,  \mu^+\,  \mu^-$ below scale $\mu_b$ have been finished
in \cite{Beneke:2017vpq,Beneke:2019slt}.

As far as  the $\mathcal{O}\left(\alpha_{\mathrm{em}}\right)$ term in 
Eq.(\ref{alphaem0}) is concerned, M. Benenke et al. found that QED virtual 
photon exchanged between one of the final-state leptons and the light 
spectator antiquark $\bar q$ in the $\bar B_q$ meson could effectively 
probe the $\bar B_q$ meson structure, resulting in a ``non-local annihilation" 
effect for muon leptonic $\bar B_q$ meson decays 
\cite{Beneke:2017vpq,Beneke:2019slt, Feldmann:2022ixt}. The spectator-$b$-quark annihilation
over the distance $1 / \sqrt{m_{B} \, \Lambda_{\mathrm{QCD}}}$ inside the 
$\bar B_q$ meson causes the strong interaction effects no longer to be 
parameterized by $f_B$ alone and provides approximately $1\%$ 
power-enhancement on branching ratio of $B_{q} \, \to\,  \mu^+\,  \mu^-$. 
This power-enhanced QED effect is substantially large and is in fact of the 
same order as the non-parametric theoretical uncertainty (about $1.5\%$
\cite{Bobeth:2013uxa}). Eventually, the theoretical uncertainty of the prediction 
on the decay $B_{s}\, \to\,  \mu^+ \, \mu^-$ is reduced largely, which will be 
necessary for matching the experimental accuracy with higher experimental 
statistics by LHCb and Belle II in the future. Recently, in \cite{Cornella:2022ubo},
M. Neubert et al. considered the virtual QED corrections though the process 
$B^-\, \to\, \mu^-\, \bar{\nu}_\mu$ to further probe the internal structure of the 
$B$-meson at subleading power in $\Lambda_{\mathrm{QCD}}/m_B$.

In view of the novel QED effect on $B_{q} \, \to \, \mu^+\,  \mu^-$ below 
$\mu_b$ scale, it would be desirable to study the other leptonic final states 
$\ell = e,\, \tau$ as QED corrections on these decays below $\mu_b$ scale 
will be process dependence. The muon mass is numerically of the order of 
the strong interaction scale $\Lambda_{\mathrm{QCD}}$, while the much 
smaller electron mass, and especially the much larger mass of the tau lepton 
imply that the results of $B_{q}\,  \to \, e^+ \, e^-$ and 
$B_{q} \, \to\,  \tau^+ \, \tau^-$ are not just trival generations from the case  
$\ell =\mu$ discussed above. In this work, we will focus on $\tau$ leptonic 
final states, $B_{q} \, \to \, \tau^+\,  \tau^-$. As the branching ratio depends 
strongly on lepton mass due to helicity suppression, tau leptonic $B$-meson 
decay is expected to have the largest leptonic branching fraction. However, 
the experimental picture for the tau channel is complicated. The necessity to 
reconstruct the tau lepton from its decay products in the presence of two or 
three undetectable neutrinos make the background rejection an experimental 
challenge. The modes of $B_{q}\, \to\,  \tau^+ \, \tau^-$ have not yet been 
experimentally observed to date. The measurements of 
$B_{q} \, \to\,  \tau^+\,  \tau^-$ at LHCb yield an upper limit for their branching ratios, 
$\mathcal{B}\left(B^{0} \, \rightarrow\, \tau^{+} \, \tau^{-}\right)\, <\, 2.1\, \times\, 10^{-3}$
\cite{LHCb:2017myy} and 
$\mathcal{B}\left(B_{s}^{0}\, \rightarrow\, \tau^{+} \, \tau^{-}\right)\, <\, 5\, \times\, 10^{-4}$
\cite{LHCb:2018roe}. 
Nevertheless, they are expected to be improved by experiments, such as 
Belle II \cite{Belle-II:2018jsg} and LHCb Upgrade II~\cite{LHCb:2018roe}, 
within the next few years. On the theoretical side, we will present an extension 
of the previous formulation in the context of SCET for muon leptonic $B$-meson 
decays to tau final states. A new element is the appearance of the order of 
hard-collinear scale $m_b\,  \Lambda_{\mathrm{QCD}}$ 
($m_{\tau}^{2} \, \sim\, m_{b}\, \Lambda_{\mathrm{QCD}}$) in the final states, 
which will be integrated out and the $\tau$ leptonic field then becomes a soft-collinear field
described by boosted heavy lepton effective theory (bHLET), similarly to the boosted HQET \cite{Fleming:2007qr,Fleming:2007xt},
not a collinear mode in $\mathrm{SCET_{II}}$ 
as for muon final state. 
It naturally makes the applications of SCET for $B_{q}\,  \to\,  \tau^+\,  \tau^-$
different from the case of $B_{q}\,  \to\,  \mu^+\,  \mu^-$. We will do two-step 
matching starting from 
$\mathrm{QCD} \times \mathrm{QED}$ onto $\mathrm{SCET_{I}}$, and 
successively onto $\mathrm{HQET} \times \mathrm{bHLET}$, 
rather than $\mathrm{SCET_{II}}$ as in muon case. Hard-collinear functions 
derived from the matching, 
$\mathrm{SCET_{I}}\, \to\, \mathrm{HQET} \times \mathrm{bHLET}$, 
will be formally response to the power enhancement term 
$m_b/\Lambda_{\mathrm{QCD}}$. However the logarithm term arising 
from the contributions of hard-collinear photon and lepton virtualities in 
the hard-collinear functions is not large for $B_{q}\, \to\, \tau^+\, \tau^-$ 
as the tau mass is just the order of hard-collinear scale, which would 
not lead to a large enhanced QED effect even though the same power 
enhancement by a factor $m_b/\Lambda_{\mathrm{QCD}}$ appears in 
$B_{q} \, \to \, \tau^+ \, \tau^-$. In addition to the hard-collinear corrections, 
one-loop hard functions will also been extracted in the first-step matching, 
$\mathrm{QCD} \times \mathrm{QED} \to \mathrm{SCET_{I}}$, for a 
complete QED correction to leptonic $B$-meson decays. 
At last, the renormalizations of $B_{q} \, \to\,  \tau^+\,  \tau^-$ will be a simple 
generation from $B_{q}\,  \to\,  \mu^+\,  \mu^-$.

The remainder of this paper is organized as follows: 
In Sec. \ref{WeakInteractions}, we briefly introduce the conventions for 
effective weak interactions for $b \, \rightarrow \, q \, \ell^{+}\,  \ell^{-}$. 
The fields and their power counting relevant to $B_{q}\, \to\,  \tau^+\,  \tau^- $ 
are discussed in Sec. \ref{power counting}. We detail the decoupling of 
hard virtualities in $\mathrm{SCET_{I}}$ and further the one of hard-collinear 
virtualities in $\mathrm{HQET} \times \mathrm{bHLET}$ in Sec. \ref{SCETI} 
and Sec. \ref{HQETSCETI}, respectively. Successively, in Sec. \ref{softmatrix}, 
matrix element of soft function in $\mathrm{HQET} \times \mathrm{bHLET}$ 
is presented. The RG evolutions involving hard functions and soft functions are 
left to Sec. \ref{RGinvolution}. The decay widths of $B_{q}\, \to\, \tau^+\, \tau^-$ 
together with the ultrasoft parts are given in Sec. \ref{ultrasoft}. We proceed with 
the numerical impact of QED corrections to $B_{q}\, \to\,  \tau^+\,  \tau^-$ in 
Sec. \ref{numerical}. Eventually, we summarize in Sec. \ref{Summary}. 
 
 \section{Effective Weak Interactions for $b \,  \rightarrow q \, \ell^{+} \,  \ell^{-}$}
 \label{WeakInteractions}
 
We start by discussing briefly the effective weak interactions for 
$b \, \rightarrow q\, \ell^{+}\, \ell^{-}$, with $\ell =e\, , \mu\, , \tau$. They can 
be firstly derived from the SM by decoupling the top quark, the Higgs boson, 
and the heavy electroweak bosons W and Z. Then the operator product 
expansion (OPE) for this effective Lagrangian relevant for $|\Delta B|\, =\, 1$ 
decays $b\, \rightarrow \, q\, \ell^{+} \, \ell^{-}$ with $q=d,\,  s$ reads
  \begin{align} \label{WeakEFT}
  \mathcal{L}_{\Delta B=1}\, =\, 
  \mathcal{N}_{\Delta B=1}\, \left[\, \sum_{i=1}^{10}\,  C_{i}\left(\mu_{b}\right)\, Q_{i}\, 
  +\, \frac{V_{u b}\,  V_{u q}^{*}}{V_{t b}\,  V_{t q}^{*}}\, \sum_{i=1}^{2} \, 
  C_{i}\left(\mu_{b}\right)\, \left(Q_{i}^{c}\, -\, Q_{i}^{u}\right)\, \right]\, 
  +\, \mathrm{h.c.}\, ,
  \end{align}
where the effective operators $Q_{i}$ are current-current operators 
$(i\, =\, 1,\, 2)$, QCD-penguin operators $(i\, =\, $ $\, 3,\,  \ldots, \, 6)$, 
dipole operators $(i\, =\, 7,\, 8)$ and semileptonic operators $(i\, =\, 9,\, 10)$. 
Here we only list those of the three most relevant operators, which followed 
the operator definitions of Ref. \cite{Chetyrkin:1996vx}, 
  \begin{align}\label{operatorQ7}
  Q_{7} &\, =\, \frac{e}{(4\,  \pi)^{2}}\, \overline{m}_{b}\,
   \left[\, \bar{q} \, \sigma^{\mu \nu}\,  P_{R} \, b\, \right]\, F_{\mu \nu} \, , \\
  Q_{9} &\, =\, \frac{\alpha_{\mathrm{em}}}{4\,  \pi}\, 
  \left(\bar{q} \, \gamma^{\mu} \, P_{L} \, b\, \right)\,
  \sum_{\ell} \, \bar{\ell}\,  \gamma_{\mu}\,  \ell \, ,\\
  Q_{10}&\, =\, \frac{\alpha_{\mathrm{em}}}{4 \, \pi}\,
  \left(\bar{q} \, \gamma^{\mu}\,  P_{L}\,  b\right)\,
  \sum_{\ell}\, \bar{\ell}\,  \gamma_{\mu}\,  \gamma_{5}\,  \ell \, ,
  \end{align}
where $\overline{m}_{b}$ represents the running $b$-quark mass in the 
$\overline{\mathrm{MS}}$ subtraction scheme. The normalization constant, 
$\mathcal{N}_{\Delta B=1}\,  \equiv \, 2 \, \sqrt{2}\, G_{F}\,  V_{t b}\,  V_{t s}^{*}$, 
is given in terms of the Fermi constant and the Cabibbo-Kobayashi-Maskawa 
(CKM) matrix elements. $C_{i}\left(\mu_{b}\right)$ denotes the 
$\overline{\mathrm{MS}}$-renormalized Wilson coefficient at the scale 
$\mu_b \, \sim \, m_b$. The matching coefficients of all of those operators 
at the electroweak scale $\mu_{W}\,  \sim\,  m_{W}$ of the order of the 
$W$-boson mass have been up to the precise of NNLO in QCD
\cite{Hermann:2013kca,Bobeth:1999mk} and further $C_{10}$ 
includes NLO EW corrections \cite{Bobeth:2013tba}. The scale running 
of $C_{i}\left(\mu\right)$ from the scale $\mu_W$ to $\mu_b$ has been 
taken into account in \cite{Bobeth:2013tba,Bobeth:2003at,Huber:2005ig,Huber:2019iqf,Huber:2020vup}, 
and the numerical values of $C_{i}\left(\mu_b\right)$ will be given in 
Section \ref{numerical}.

 \section{Factorization in $B_{q} \, \rightarrow \, \ell^{+} \, \ell^{-}$ decay 
               below the scale $m_b$}
 \label{Sec3}
 
The heavy-quark systems $B_{q}$ can be described well by heavy-quark 
effective theory (HQET) \cite{Georgi:1990um}. The process 
$B_{q}\, \rightarrow\, \ell^{+} \, \ell^{-}$ also involves final energetic light 
particles where some components of their momentas $p_\mu$ are large, 
but their $p^2$ are small when compared with the heavy $B$-meson. 
More specifically, working in the rest frame of the initial $B$-meson 
and choosing the $z$-direction as the direction of the one of the two
leptons, their momentas can be written as
  \begin{align}
  \begin{aligned}
  p_{\ell^+}^\mu&\, =\, (E_{\ell^+},\, 0,\, 0,\, \sqrt{m_B^2\, -\, 4\,  m_\ell^2}/2)\, ,\\
  p_{\ell^-}^\mu&\, =\, (E_{\ell^-},\, 0,\, 0,\, -\sqrt{m_B^2\, -\, 4 \, m_\ell^2}/2) \, ,
  \end{aligned}
  \end{align}
where the large energies are $E_{\ell^+}\, =\, E_{\ell^-}\, =\, m_B/2$ 
and the final-state leptons are on-shell, 
$p_{\ell^+}^2\, =\, p_{\ell^-}^2\, =\, m_\ell^2$. The presence of 
several different scales in $B_{q}\, \rightarrow\, \ell^{+}\, \ell^{-}$ 
decay means that we can classify quantum fluctuations as hard, 
hard-collinear (collinear), or soft. For $\ell\, =\, \tau$, the 
corresponding scales are
  \begin{align}
  \begin{aligned}
  \text { hard: } & m_{B}\, , E_{\tau}\, , \\ 
  \text { hard-collinear: } & m_{\tau} \, \sim \,  
           \sqrt{m_{B}\, \Lambda_{\mathrm{QCD}}} \, , \\ 
  \text { soft: } & \Lambda_{\mathrm{QCD}} \, .
  \end{aligned}
  \end{align}
Our goal is to integrate out all short-distance scales including hard 
and hard-collinear quantum fluctuations. Therefore the construction 
of EFTs often proceeds two-step matching procedure: in the first step, 
hard quantum fluctuations are integrated out by matching the effective 
weak Lagrangian in Eq.(\ref{WeakEFT}) onto $\mathrm{SCET_{I}}$ 
with hard-collinear or soft momenta as dynamical degrees of freedom; 
in the second step, by matching $\mathrm{SCET_{I}}$ onto 
$\mathrm{HQET} \times \mathrm{bHLET}$, fluctuations at the 
hard-collinear scale are integrated out. The explicit factorizations of 
the two short-distance scales from long-distance scale will be done 
in the following two subsections.

\subsection{Power Counting}
\label{power counting}

In view of the presence of fast, hard-collinear final particles, it is 
convenient to decompose 4-vectors in a light-cone basis spanned 
by two light-like reference vectors $n_+^{\mu}$, $n_-^{\mu}$ and 
a remainder perpendicular to both. We often choose 
$n_+^{\mu}\, =\, (1,\, 0,\, 0,\, 1)$ and $n_-^{\mu}\, =\, (1,\, 0,\, 0,\, -1)$ 
to make one of the two final states align along the $n_+^\mu$ 
direction, and the other point the opposite direction, $n_-^{\mu}$. 
An arbitrary vector $p^{\mu}$ can then be decomposed in a 
component proportional to $n_+^{\mu}$, a part proportional to 
$n_-^{\mu}$, and the transverse direction,
  \begin{align}
  \begin{aligned}
  p^{\mu}&\, =\, 
  (n_+p)\, \frac{n_-^{\mu}}{2}\, +\, (n_-p)\, \frac{n_+^{\mu}}{2}\, 
   +\, p_{\perp}^{\mu}\\
  & \, \equiv\, \left(n_+ p\, ,  \, n_-p\, ,  \, p_{\perp}\right)\, .
  \end{aligned}
  \end{align}
On the partonic level, $B_{q}\,  \rightarrow\,  \ell^+\, \ell^-$ 
decay processes as
  \begin{align}
  b\left(p_{b}\right)\, +\, q\left(l_{q}\right) \, \rightarrow \, 
  \ell^+ \left(p_{\ell^+}\right)\, +\, \ell^- \left(p_{\ell^- }\right)\, .
  \end{align}
The momentums of two final states are decomposed as
  \begin{align}
  p_{\ell^+}^{\mu}&\, =\,  
  \frac{m_B\, -\, \sqrt {m_B^2\, -\, 4\, m_{\ell}^2}}{2}\, \frac{n_-^{\mu}}{2}\, +\, 
  \frac{m_B\, +\, \sqrt {m_B^2\, -\, 4\, m_{\ell}^2}}{2}\, \frac{n_+^{\mu}}{2}\, ,\\
  p_{\ell^-}^{\mu}&\, =\, 
  \frac{m_B\, +\, \sqrt {m_B^2\, -\, 4\, m_{\ell}^2}}{2} \, \frac{n_-^{\mu}}{2}\, +\, 
  \frac{m_B\, -\, \sqrt {m_B^2\, -\, 4\, m_{\ell}^2}}{2}\, \, \frac{n_+^{\mu}}{2}\, .
  \end{align}
Specifically, for $\ell \, =\,  \tau$, $n_+\, p_{\tau^-}\, =\, n_-\, p_{ \tau^+} \, 
\sim\,  m_b$ and $n_-\, p_{\tau^-}\,  =\, n_+\, p_{\tau^+} \, \sim\,  
\Lambda_{\mathrm{QCD}}$. A softly interacting heavy $b$-quark 
is nearly on-shell with its momentum 
$p^{\mu}_{b}\, =\, m_{b}\,  v^{\mu}\, +\, l_{b}^{\mu}\, $, where $v^\mu$ 
is the 4-velocity of the $B_q$ meson, $v^\mu = (n_+^\mu + n_-^\mu)/2$,  
and the ``residual momentum" $l_{b}\,\sim \Lambda_{\mathrm{QCD}}$.
Also the momentum of light spectator quark is 
$l_{q} \, \sim \, \Lambda_{\mathrm{QCD}}$.
 
Besides the external kinematics above for 
$B_{q}\, \rightarrow\, \tau^+\, \tau^-$ decay, the internal dynamic 
momentum, denoted by $k^\mu$, can be classified according to 
their scaling properties with $m_b\, \gg\, \Lambda_{\mathrm{QCD}}$ 
as
  \begin{align} 
  \begin{aligned} 
  \text { hard: } &\,  k_{h}^{\mu} \, = \,  m_b\, (1,\, 1,\, 1) \sim(1,\, 1,\, 1)\, , \\
  \text { hard-collinear: } &\,  k_{h c}^{\mu}  \, = \, 
   (m_{b},\,  \Lambda_{\mathrm{QCD}},\, \sqrt{m_{b}\,  \Lambda_{\mathrm{QCD}}})
  \, \sim\,  (1, \, \lambda^{2}, \, \lambda)\, , \\
  \text {anti-hard-collinear: } &\,  k_{\overline{hc}}^{\mu}\, =\, 
   (\Lambda_{\mathrm{QCD}},\, m_{b},\, \sqrt{m_{b}\,  \Lambda_{\mathrm{QCD}}})
  \, \sim \, ( \lambda^{2},\, 1, \, \lambda) \, , \\
  \text { soft: } &\,  k_{s}^{\mu}\, =\,  
  (\Lambda_{\mathrm{QCD}}, \, \Lambda_{\mathrm{QCD}}, \, \Lambda_{\mathrm{QCD}})
  \, \sim\, (\lambda^{2}, \, \lambda^{2}, \, \lambda^{2})\, ,\\
  \text { soft-collinear: } &\,  k_{sc}^{\mu}\, =\,  
 (1/b, \,b , \, 1)\, \Lambda_{\mathrm{QCD}},
  \, \sim\, (1/b, \,b , \, 1)\, \lambda^{2}\, ,\\
    \text { anti-soft-collinear: } &\,  k_{sc}^{\mu}\, =\,  
 (b , \,1/b, \, 1)\, \Lambda_{\mathrm{QCD}},
  \, \sim\, (b , \, 1/b, \,1)\, \lambda^{2}\, ,
  \end{aligned}
  \end{align}
with scaling parameter $\lambda^{2}\, =\, \Lambda_{\mathrm{QCD}}/ m_{b}$ 
and the boosted parameter $b = m_\tau/m_b$.
The corresponding virtualities are $k_{h}^2\, \sim\, m_b^2 $, 
$k_{hc}^2\,= \, k_{\overline{hc}}^2\, \sim\, m_{b} \, \Lambda_{\mathrm{QCD}} $, 
and $k_{s}^2\, \sim\, \Lambda_{\mathrm{QCD}}^{2}$. Different from the light 
final particles $\ell\, =\, \mu$, collinear virtuality
 $k_{c}^2\, \sim\, \Lambda_{\mathrm{QCD}}^{2}$ does not appear in massive 
 $\tau$ lepton case ($m_{\tau}^{2}\, \sim\, m_{b}\, \Lambda_{\mathrm{QCD}}$). 
 The massive $\tau$ field will be integrated out and become a soft-collinear field in bHLET.
Consequently, the matching procedure of EFTs would also be different from $B_{d,s} \to \mu^+ \mu^-$. 
As mentioned above, after integrating out hard modes (i.e., $k_{h}^2\, \sim\, m_b^2 $), 
we obtain the $\mathrm{SCET_I}$ including the (anti-)hard-collinear and soft (soft-collinear) modes. 
Subsequently, the (anti-)hard-collinear modes of light quark 
(i.e., $k_{hc}^2\,= \, k_{\overline{hc}}^2\, \sim\, m_{b}\, \Lambda_{\mathrm{QCD}} $) 
will be integrated out to become the soft field in HQET, but the (anti-)hard-collinear $\tau$ field 
will be turned to be the (anti-)soft-collinear field after integrating out the mass of $\tau$ lepton in bHLET.
The matching procedure simply follows as
  \begin{align*}
  \hskip0.5cm \text{full QED}\hskip0.7cm & & & 
  \to & \text{\SCETI{}} & & & 
  \hskip -0.6cm  \to \hskip 1.5cm \text{$\mathrm{HQET} \times \mathrm{bHLET}$}
  \\
  \text{hard:} \; \mu_h^2 \sim m_b^2 & & & & 
  \text{hard-collinear:}  & \; \mu_{hc}^2 \sim m_b  \LambdaQCD & & 
  \text{soft(-collinear):} \; \mu_{s (c)}^2 \sim  \LambdaQCD^2 
  \end{align*}
At last, we introduce various fields of $\text{\SCETI}$ and $\mathrm{HQET} \times \mathrm{bHLET}$ 
obtained by decomposing the quark, lepton and  photon (gluon) fields into 
various momentum modes. The fields and their scalings are
  \begin{align}
  \begin{aligned} 
  \label{filedscale}
  \text {soft heavy quark:} & \quad h_{v} \, \sim\,  \lambda^{3}\, ,\\
  \text {hard-collinear light quark:}& \quad \chi_{hc}\,  \sim\,  \lambda  \, ,\\
  \text {hard-collinear leptonic field:}& \quad \ell_{hc}\,  \sim\,  \lambda  \, ,\\
  \text {soft light quark:}& \quad q_{s}\,  \sim\,  \lambda^{3} \, ,\\
  \text {soft-collinear leptonic field:}& \quad \ell_{sc}\,  \sim\,  \lambda^{3} \, ,\\
  \text {hard-collinear photon\,  (gluon):}& \quad A_{h c}^{\mu} (G_{h c}^{\mu}) 
   \, \sim\,  (1,\, \lambda^2, \, \lambda)\, ,\\
  \text {soft photon\, (gluon):} &\quad A_{s}^{\mu}(G_{s}^{\mu}) 
   \, \sim\,  \lambda^2\, (1,\, 1,\, 1)\, .
  \end{aligned}
  \end{align}

\subsection{$\mathrm{SCET_I}$}
 \label{SCETI}
 
In this subsection, we will present the effective operators in 
$\mathrm{SCET_I}$ and hard fluctuations decoupled in the 
matching of weak EFT onto $\mathrm{SCET_I}$ up to NLO.

\subsubsection[Operators]
 {\boldmath Operators}
 
After introducing the relevant fields and discussing their power 
counting, we proceeded to present $\mathrm{SCET_I}$ 
operators in the matching of the effective weak Hamiltonian to 
$\mathrm{SCET_I}$. As the $\mathrm{SCET_I}$ operators for 
$B_q \, \to \, \tau^+\,  \tau^-$ decay are the same as in $\mu$ 
leptonic decay \cite{Beneke:2019slt}, we just list those operators 
here and the details of their constructions can be found in the 
Appendix of that paper \cite{Beneke:2019slt} and earlier works
\cite{Beneke:2003pa,Beneke:2017ztn}. With the power counting 
of fields in Eq.(\ref{filedscale}), the smallest $\lambda$ scaling of 
$\mathrm{SCET_I}$ operators relevant to the matching of effective 
operators $Q_{9,10}$ are order of $\lambda^6$. 
In the coordinate-space, labelled by a tilde, they are
  \begin{align}
  \widetilde{\mathcal{O}}_{9}(s, t) &\, =\,g_{\mu \nu}^{\perp}\,
  \left[\, \bar{\chi}_{hc}\left(s n_{+}\right)\, \gamma_{\perp}^{\mu}\, 
        P_{L}\, h_{v}(0)\, \right]\,
  \left[\, \bar{\ell}_{hc}\left(t n_{+}\right)\, \gamma_{\perp}^{\nu}\, 
        \ell_{\overline{hc}}(0)\, \right] \, ,\\
  \widetilde{\mathcal{O}}_{10}(s, t) &\, =\, i \varepsilon_{\mu \nu}^{\perp}\,
  \left[\, \bar{\chi}_{hc}\left(s n_{+}\right)\, \gamma_{\perp}^{\mu}\, 
        P_{L}\, h_{v}(0)\, \right]
  \left[\, \bar{\ell}_{hc}\left(t n_{+}\right)\, \gamma_{\perp}^{\nu}\, 
         \ell_{\overline{hc}}(0)\, \right] \, ,
  \end{align}
for a hard-collinear light quark, that is, the light quark is parallel to 
the hard-collinear lepton, and 
  \begin{align}
  \widetilde{\mathcal{O}}_{\overline{9}}(s, t) &\, =\, g_{\mu \nu}^{\perp}\, 
  \left[\, \bar{\chi}_{\overline{hc}}\left(s n_{-}\right)\,  \gamma_{\perp}^{\mu}\,  
         P_{L}\,  h_{v}(0)\, \right]\, 
  \left[\, \bar{\ell}_{hc}(0)\,  \gamma_{\perp}^{\nu} \, 
         \ell_{\overline{hc}}\left(t n_{-}\right)\, \right] \, ,\\ 
  \widetilde{\mathcal{O}}_{\overline{10}}(s, t) &\, =\, i \varepsilon_{\mu \nu}^{\perp}\, 
  \left[\, \bar{\chi}_{\overline{hc}}\left(s n_{-}\right)\, \gamma_{\perp}^{\mu}\,  
         P_{L}\, h_{v}(0)\, \right]
  \left[\, \bar{\ell}_{hc}(0)\,  \gamma_{\perp}^{\nu}\, 
         \ell_{\overline{hc}}\left(t n_{-}\right)\, \right]\, , 
  \end{align}
for an anti-hard-collinear quark, respectively. Tensors 
$g_{\mu \nu}^{\perp}$ and $\varepsilon_{\mu \nu}^{\perp}$ 
are defined as $g_{\mu \nu}^{\perp}\, \equiv\,  g_{\mu \nu}\, -\, 
\frac{n_{+}^{\mu}\, n_{-}^{\nu}}{2}\, -\, \frac{n_{-}^{\mu}\, n_{+}^{\nu}}{2}, \, \,
\varepsilon_{\mu \nu}^{\perp} \, \equiv\,  
\varepsilon_{\mu \nu \alpha \beta}\, \frac{n_{+}^{\alpha} \, n_{-}^{\beta}}{2}$.
Actually, using the formula 
$i\, \varepsilon_{\mu\nu}^\perp\, \gamma_\perp^\nu\, =\, 
\gamma_{\perp\, \mu}\, \gamma_5$ established in dimension 
$d=4$, we find that operators $ \widetilde{\mathcal{O}}_{9}(s, t)$ 
and $ \widetilde{\mathcal{O}}_{10}(s, t)$,
$\widetilde{\mathcal{O}}_{\overline{9}}(s, t)$ and 
$\widetilde{\mathcal{O}}_{\overline{10}}(s, t)$ are 
not independent, and their relations are
  \begin{align}\label{910relation}
  \widetilde{\mathcal{O}}_{9}(s, t)\,  =\,  
  \widetilde{\mathcal{O}}_{10}(s, t)\, , \quad 
  \widetilde{\mathcal{O}}_{\overline{9}}(s, t)\, =\,
  -\, \widetilde{\mathcal{O}}_{\overline{10}}(s, t)\, .
  \end{align}
We usually do matching in momentum space, and 
$\widetilde{\mathcal{O}}_{9\, (\bar 9)}(s, t)$ can be 
Fourier transformed to $\mathcal{O}_{9 \, (\bar 9)}(u)$ 
as
  \begin{eqnarray}
  \mathcal{O}_{9}(u)\, =\, 
  n_{+} p_{hc} \, \int\,  \frac{d r}{2 \pi}\, e^{-i \, u\,  r\, \left(n_{+} p_{hc}\right)}\, 
  \widetilde{\mathcal{O}}_{i}(0, r)\, ,
  \end{eqnarray}
where only single variable $u$ is introduced once we use the
hard-collinear momentum conservation with the total hard-collinear 
momentum $n_{+} p_{hc}\, =\, n_{+}\left(p_{\chi}\, +\, p_{\ell}\right)$ 
and it should be interpreted as the fraction $n_{+} p_{\ell} / n_{+} p_{hc}$ 
of $n_{+} p_{hc}$ carried by one of two lepton fields, and then $\bar{u}$
is the momentum fraction of  hard-collinear light quark in $B$-meson, 
$\bar{u} \, \equiv\, (1\, -\, u)\, =\, n_{+} p_{\chi} / n_{+} p_{hc}$.  
The operator $\widetilde{\mathcal{O}}_{\bar{9}}$ can be defined 
similarly by replacing $n_{+}$ by $n_{-}$.

The weak EFT operator $Q_7$ in Eq.(\ref{operatorQ7}) can 
also be matched to $\mathcal{O}_9$ in $\mathrm{SCET_I}$ 
by integrating out hard photon from the electromagnetic 
dipole operator,
  \begin{align}
  Q_{7}\, =\, \frac{2 \, Q_{\ell}}{u}\,  \mathcal{O}_{9}\, ,
  \end{align}
where $Q_{\ell}\, =\, -1$.
 
It can be seen from above that, in four-dimensional space-time 
($d \, =\, 4$), only $\mathcal{O}_{9,\,  {\bar 9}}$ are physical 
operators in $\mathrm{SCET_I}$. In fact, a complete basis 
should also contain evanescent operators when we use 
dimensional regularization $(d \, =\, 4\, -\, \epsilon)$ in 
calculating the matrix elements of operators. The generic 
feature of these evanescent operators is that they vanish 
after going to four-dimensional space-time, $d \, \to\,  4$. 
More precisely, because $\widetilde{\mathcal{O}}_{9}$ 
tends to be equal to $\widetilde{\mathcal{O}}_{10}$ in 
$d \, \to\,  4$, that is,
  \begin{align}
  \widetilde{\mathcal{O}}_{9}\, 
  \stackrel{\mathrm{d=4}}{\longrightarrow} \, \widetilde{\mathcal{O}}_{10}\, , 
  \end{align}
the evanescent operator, denoted by $\widetilde{\mathcal{O}}_{E}$, 
can be defined as
  \begin{align}
  \begin{aligned}
  \widetilde{\mathcal{O}}_{E}(s,t) &\, \equiv\, 
  \widetilde{\mathcal{O}}_{9}(s,t)\, -\, \widetilde{\mathcal{O}}_{10}(s,t) \\
  &\, =\, \frac{1}{2}\,
   \left[\, \bar{\chi}_{hc}\left(s n_{+}\right)\,  \gamma_{\perp}^{\mu}\,  
          P_{L}\,  h_{v}(0)\, \right]
  \left[\, \bar{\ell}_{hc}\left(t n_{+}\right)\,  \gamma^{\perp}_{\mu}\, 
          P_L\, \ell_{\overline{hc}}(0)\, \right] \, .
  \end{aligned}
  \end{align}
More evanescent operators will appear when we do 
matching from $\mathrm{QCD} \times \mathrm{QED}$ 
to $\mathrm{SCET_I}$ to higher order and their 
definitions will be given specifically until then.

\subsubsection{Matching from $\mathrm{QCD} \times \mathrm{QED}$ 
                         to $\mathrm{SCET_I}$}
  
We will integrate out hard fields by matching effective operators 
$Q_k\, , k=1,\, ...\, ,6,\, 7,\,  9,\, 10,$ in 
$\mathrm{QCD} \times \mathrm{QED}$ onto 
$\mathcal{O}_{9}$ and $\mathcal{O}_{E}$ in $\mathrm{SCET_I}$. 
The hard matching condition at the scale $\mu_{b}$ is given by
  \begin{figure}[h]
  \begin{center}
  \vspace{0cm}\hskip0cm
  \includegraphics[width=0.5\textwidth]{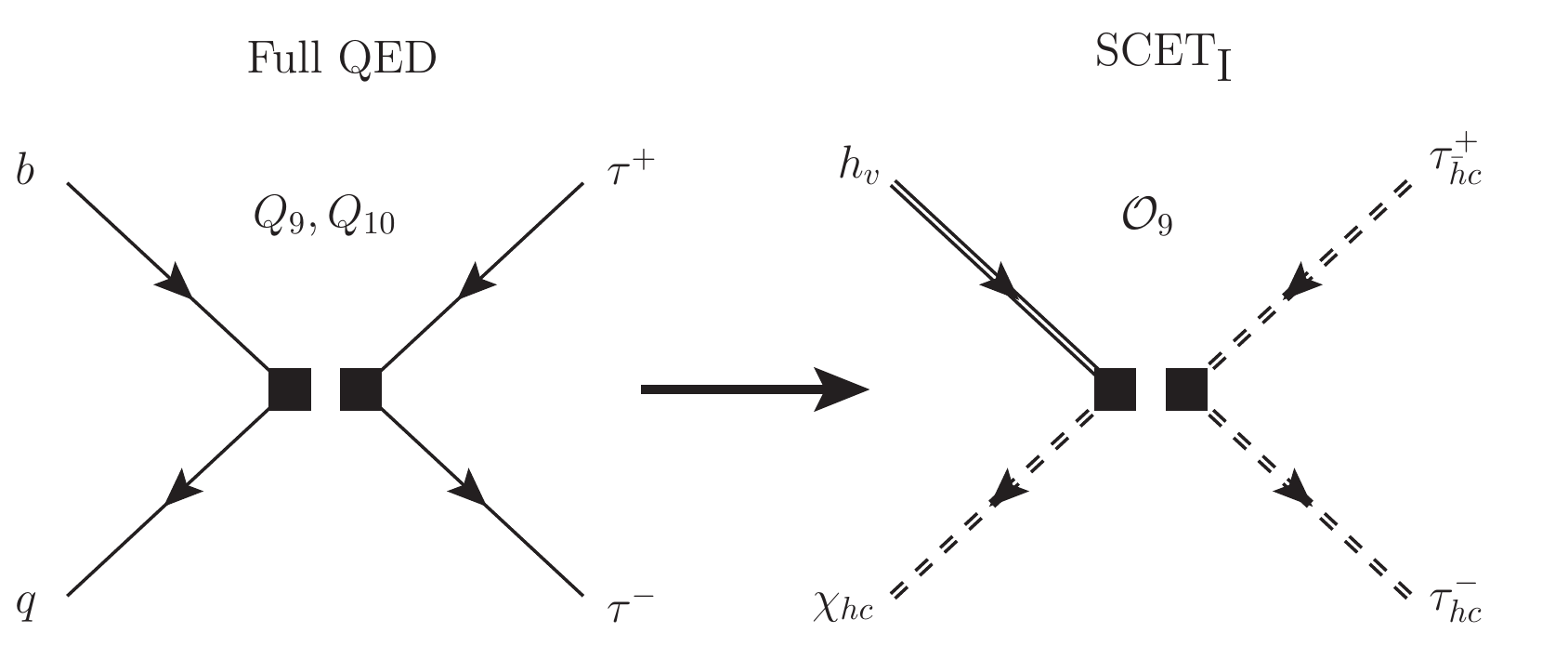}
  \end{center}
  \caption{
  The figure shows the tree-level matching of $\mathrm{QED}$ 
  onto $\mathrm{SCET_I}$, with EFT operators $Q_{9,\, 10}$ 
  and $\mathrm{SCET_I}$ operator $\mathcal{O}_{9}$ contributing, 
  respectively. The notation and scaling of the $\mathrm{SCET_I}$ 
  fields are given in Eq.(\ref{filedscale}). The heavy quark field $h_v$ 
  in HQET is labelled by a double-solid line and (anti-)hard-collinear 
  fields $\xi_{hc}$, $\tau_{hc}$ in $\mathrm{SCET_I}$ are denoted 
  by double-dashed lines.}
  \label{fig:QEDSCETmatching}
  \end{figure}

  \begin{eqnarray}
  \mathcal{N}_{\Delta B=1} \, \sum_{k}\,  C_{k}\, \left(\mu_{b}\right) Q_{k}\, 
  =\, \sum_{i} \, \int d u\, H_{i}\left(u, \mu_{b}\right)\,  \mathcal{O}_{i}(u)\, ,
  \end{eqnarray}
where $i\, =\, 9, \, E$. $C_{k}$ is the Wilson coefficient and 
$H_{i}$ represents hard function. By calculating the appropriate 
matrix elements in both sides of the equation above up to NLO 
in the order of $\alpha_{\mathrm{em}}$, firstly we get the hard 
functions at tree-level,
  \begin{align}
  H_{9}^{(0)}\left(u, \mu_{b}\right)&\, =\, \mathcal{N}\, 
  \left[\, C_{9}^{(0)} \left(u, \mu_{b}\right)\, +\, C_{10}^{(0)} (\mu_{b})\, 
  -\, \frac{2 Q_{\ell}}{u}\,  C_{7}^{(0)}\left(u, \mu_{b}\right)\, \right]\, ,\\
  H_{E}^{(0)}\left(u, \mu_{b}\right)&\, =\, \mathcal{N}\, 
  \left[\, - C_{10}^{(0)} (\mu_{b})\, \right]\, ,
  \end{align}
with 
  \begin{eqnarray}
  \mathcal{N}\,  \equiv\,  \mathcal{N}_{\Delta B=1} \, 
  \frac{\alpha_{\mathrm{em}}\left(\mu_{b}\right)}{4 \pi}\, ,
  \end{eqnarray}
where $C_{7\, ,\, 9\, ,\, 10}^{(0)} $ are the Wilson coefficients of 
$Q_{7\, ,\, 9\, ,\, 10}$ at LO in $\alpha_{\mathrm{em}}$. Formally, 
the contribution from four-quark operators $Q_{k}(k\, =\, 1, \ldots, 6)$ 
will start from one loop level. However these quark loops can be fully 
absorbed into effective Wilson coefficients $C_{7,\, 9}^{\mathrm{eff}}$
\cite{Buras:1994dj}, and then the hard function 
$H_{9}^{(0)}\left(u, \mu_{b}\right)$ decoupled from all four-quark 
operators $Q_k\, , k=1,\, ...\, ,6,\, 7,\,  9,\, 10,$ at tree level should 
be replaced by
  \begin{align}\label{H90}
  H_{9}^{(0)}\left(u, \mu_{b}\right)\, =\, \mathcal{N}\, 
  \left[\, C_{9}^{\mathrm{eff}}\left(u, \mu_{b}\right)\, +\, C_{10}^{(0)} (\mu_{b})\, 
  -\, \frac{2 \, Q_{\ell}}{u} \, C_{7}^{\mathrm{eff}}\left(u, \mu_{b}\right)\, \right]\, .
  \end{align}
For the case of an anti-hard-collinear quark, the hard function 
decoupled from $\mathcal{O}_{\bar{9}}$ is
  \begin{align}\label{H90bar}
  H_{\bar 9}^{(0)}\left(u, \mu_{b}\right)
  &\, =\, \mathcal{N}\, \left[\, C_{9}^{\mathrm{eff}}\left(u, \mu_{b}\right)\, -\, 
  C_{10}^{(0)}\left(\mu_{b}\right)\, -\, 
  \frac{2 Q_{\ell}}{u} \, C_{7}^{\mathrm{eff}}\left(u, \mu_{b}\right)\, \right]\, ,
  \end{align}
where the minus in front of $C_{10}^{(0)}\left(\mu_{b}\right)$ is 
due to the opposite relation of operator 
$\widetilde{\mathcal{O}}_{\overline{9}}\, $ and 
$\widetilde{\mathcal{O}}_{\overline{10}}\,$ in Eq.(\ref{910relation}).

We can obtain the hard functions $H^{(1)}_{i}$ by expanding 
the matching equation at NLO as 
  \begin{eqnarray}\label{HmatchingNLO}
  \mathcal{N}_{\Delta B=1} 
  \sum_{k} C^{(1)}_{k}\left(\mu_{b}\right) \langle Q_{k}\rangle^{(0)}
  =\sum_{i} \int d u \left[ H^{(0)}_{i}\left(u, \mu_{b}\right)
   \langle\mathcal{O}_{i}(u)\rangle^{(1)}+ 
   H^{(1)}_{i}\left(u, \mu_{b}\right) 
   \langle\mathcal{O}_{i}(u)\rangle^{(0)}\right],\nonumber\\
  \end{eqnarray}
with 
  \begin{align}
  \langle\, \mathcal{O}_{i}(u)\, \rangle^{(1)}\, =\, 
  Z_{i j}^{(1)}\,  \langle\, \mathcal{O}_{j}(u)\, \rangle^{(0)}\, ,
  \end{align}
in Dimension Regulation, where $Z_{i j}^{(1)}$ is the 
UV renormalization factor of $\mathcal{O}_{j}(u)$. 
The Wilson coefficient $C^{(1)}_{k}$ represents 
the one at NLO of $\alpha_{\mathrm{em}}$. 
When we calculate the matrix element of l.h.s of 
Eq.(\ref{HmatchingNLO}) at NLO, more operators 
will be involved, which are written, without the position
variables, as
  \begin{align}
  \begin{aligned} 
  {\mathcal{O}}_{9,1}&\,  \equiv\, 
  [\, \bar{\chi}_{hc}\, \gamma^{\mu}\,  \gamma^{\nu}\,  \gamma^{\rho}\, 
    P_{L}\, h_{v}\, ]\,
  [ \, \bar{\ell}_{hc}\, \gamma_{\mu}\,  \gamma_{\nu}\,  \gamma_{\rho}\, 
    \ell_{\overline{hc}}\, ] \, ,\\
  {\mathcal{O}}_{9,2}&\,  \equiv\, 
  [\, \bar{\chi}_{hc}\, \gamma^{\mu} \, \gamma^{\nu}\,  \gamma^{\rho}\, 
    P_{L}\, h_{v}\, ]\,
  [ \, \bar{\ell}_{hc}\, \gamma_{\rho} \, \gamma_{\nu}\, \gamma_{\mu}\,  
    \ell_{\overline{hc}}\, ] \, ,\\
  {\mathcal{O}}_{10,1}& \, \equiv\, 
  [\, \bar{\chi}_{hc}\, \gamma^{\mu}\,  \gamma^{\nu} \, \gamma^{\rho}\, 
    P_{L}\, h_{v}\, ]\,
  [ \, \bar{\ell}_{hc}\, \gamma_{\mu}\,  \gamma_{\nu}\,  \gamma_{\rho}\,  
    \gamma_5\, \ell_{\overline{hc}}\, ] \, ,\\
  {\mathcal{O}}_{10,2}& \, \equiv\, 
  [\, \bar{\chi}_{hc}\, \gamma^{\mu} \, \gamma^{\nu} \, \gamma^{\rho}\, 
    P_{L}\, h_{v}\, ]\,
  [\,  \bar{\ell}_{hc}\, \gamma_{\rho}\,  \gamma_{\nu}\, \gamma_{\mu}\,  
    \gamma_5\, \ell_{\overline{hc}}\, ]\, .
  \end{aligned} 
  \end{align}
It is easy to find that 
$\mathcal{O}_{9,1}\, +\, \mathcal{O}_{9,2} \, =\, 20\, \mathcal{O}_{9}$,
$\mathcal{O}_{10,1}\, +\, \mathcal{O}_{10,2} \, = \, 20\, \mathcal{O}_{10}$, 
and $\mathcal{O}_{9,1}\, =\, 4\, \mathcal{O}_{9}$, 
$\mathcal{O}_{10,1}\, =\, 4\, \mathcal{O}_{10}$ in $d=4$. 
We can possibly choose the following evanescent operators,
  \begin{align}
  \begin{aligned} 
  \mathcal{O}_{E_1}&\, \equiv\,  {\mathcal{O}}_{9,1}\, -\, 4\, {\mathcal{O}}_{9}\, ,\\
  \mathcal{O}_{E_2}&\, \equiv\,  {\mathcal{O}}_{10,1}\, -\, 4\,{\mathcal{O}}_{10} \\
  &\, =\, {\mathcal{O}}_{10,1}\, -\, 4\, \mathcal{O}_{9}\, +\, 4\, \mathcal{O}_{E}\, .
  \end{aligned} 
  \end{align}
It is clearly that the physical operator ${\mathcal{O}}_{9}$ 
and the evanescent operators 
$\mathcal{O}_{E},\, \mathcal{O}_{E_1},\, \mathcal{O}_{E_2}$ 
will be contained in $\mathrm{SCET_I}$ when we consider 
the correction to NLO.

The hard functions at NLO can be extracted as
  \begin{eqnarray}
  H^{(1)}_{i}\left(u, \mu_{b}\right)\, =\, 
  \mathcal{N}\,  C^{(1)}_{k}\left(\mu_{b}\right)\, -\, 
  H^{(0)}_{j}\left(u, \mu_{b}\right)\, Z_{ji}^{(1)}\, ,
  \end{eqnarray}
with $k=7,\, 9,\, 10\, $ and 
$i \, \, \text{or}\, \,  j=9\, , \, E\, , \, E_1\, , \, E_2$.
In the following, we concentrate on $i=9$, and the 
hard function associate with physical operator 
${\mathcal{O}}_{9}$ is
  \begin{eqnarray}\label{Hard9NLO1}
  H^{(1)}_{9}\left(u, \mu_{b}\right)=
  \mathcal{N} \left[\, C^{(1)}_{7, \, 9}\left(\mu_{b}\right)+
  C^{(1)}_{10}\left(\mu_{b}\right)\, \right] - 
  H^{(0)}_{9}\left(u, \mu_{b}\right)\, Z_{99}^{(1)}- 
  H^{(0)}_{E}\left(u, \mu_{b}\right)\, Z_{E9}^{(1)}\, ,
  \end{eqnarray}
where terms for $ j= E_1\, ,\, E_2$ disappear due to 
$H^{(0)}_{E_1,\, E_2}=0$. Next it is necessary to 
calculate the matrix element of evanescent operator 
$\mathcal{O}_{E}$ at NLO to check whether 
$\mathcal{O}_{E}$ would contribute to hard function 
$H^{(1)}_{9}$ or not, and the result is
  \begin{align}
  \langle  \mathcal{O}_{E} \, \rangle^{(1)} \, \sim \, 
  H_{E}^{(1)}\, \langle  \mathcal{O}_{E}  \, \rangle^{(0)}\, ,
  \end{align}
where the physical amplitude 
$\langle {\mathcal{O}}_{9} \, \rangle^{(0)}\,$ do not appear.
It means that evanescent operator $\mathcal{O}_{E}$ does 
not have an influence on hard function $H_{9}$ at one loop 
level and Eq.(\ref{Hard9NLO1}) can be reduced to 
  \begin{eqnarray}\label{Hard9NLO2}
  H^{(1)}_{9}\left(u, \mu_{b}\right)\, =\, 
  \mathcal{N}\, \left[\, C^{(1)}_{7, \, 9}\left(\mu_{b}\right)\, +\, 
  C^{(1)}_{10}\left(\mu_{b}\right)\, \right]\, 
  -\, H^{(0)}_{9}\left(u, \mu_{b}\right)\, Z_{99}^{(1)}\, .
  \end{eqnarray}
For the case of an anti-hard-collinear quark 
$\mathcal{O}_{\bar{9}}$, the corresponding hard function 
is
  \begin{align}\label{Hard9barNLO2}
  H_{\bar 9}^{(1)}\left(u, \mu_{b}\right)\, =\, 
  \mathcal{N}\, \left[C^{(1)}_{\bar 7, \, \bar 9}\left(\mu_{b}\right)\, -\, 
  C^{(1)}_{\overline{10}}\left(\mu_{b}\right)\right]\, -\, 
  H^{(0)}_{\bar 9}\left(u, \mu_{b}\right)\, Z_{\bar 9 \bar 9}^{(1)}\, ,
\end{align}
where the second terms in r.h.s of both Eqs.(\ref{Hard9NLO2}) 
and (\ref{Hard9barNLO2}) are the IR subtractions to cancel 
the IR divergences from their first terms.

\begin{figure}[h]
  \begin{center}
  \vspace{0cm}\hskip0cm
  \includegraphics[width=0.65\textwidth]{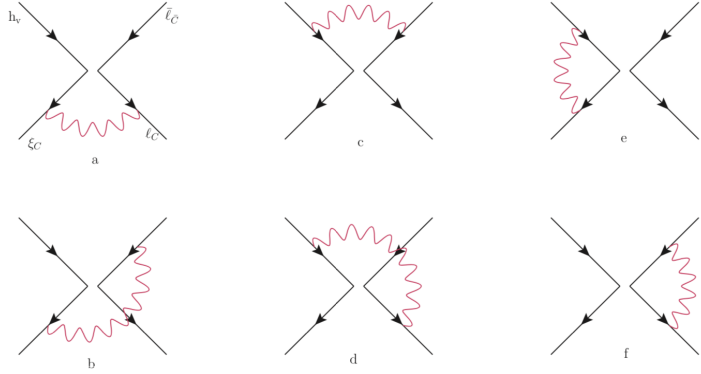}
  \end{center}
  \caption{one-loop QED corrections to hard functions}
  \label{fig:loopmatching}
  \end{figure}

The first terms in Eqs.(\ref{Hard9NLO2}) and (\ref{Hard9barNLO2}) 
are from one loop contributions with $Q_k$ and $Q_{\bar k}$,\, 
$k=7,\, 9,\, 10,$ inserted in Fig.(\ref{fig:loopmatching}), which 
are corresponding to the diagrams with a hard-collinear and 
an anti-hard-collinear light quark state, respectively.
We find that the results in the case of the hard-collinear 
sector are only equal to ones of the anti-hard-collinear 
sector for Figs.(e) and (f), while opposite for Figs.(a)-(d), 
e.g.
  \begin{align}
  \begin{aligned}\label{OneloopHard}
  C_{k}^{\text {(a)}}\,&=\, -\, C_{\bar k}^{\text {(b)}}\, ,\quad  \quad
  C_{k}^{\text {(b)}}\,=\, -\,  C_{\bar k}^{\text {(a)}}\, ,\\
  C_{k}^{\text {(c)}}\,&=\, -\, C_{\bar k}^{\text {(d)}}\, , \quad  \quad
  C_{k}^{\text {(d)}}\,=\, -\, C_{\bar k}^{\text {(c)}}\, ,\\
  C_{k}^{\text {(e)}}\,&=\, \, C_{\bar k}^{\text {(e)}}\, ,\quad  \quad \quad
  C_{k}^{\text {(f)}}\,=\, \, C_{\bar k}^{\text {(f)}}\, ,
  \end{aligned}
  \end{align}
where $C_{k}^{\text {(a-f)}}$ denotes contribution from 
hard mode integrated out from each diagram in 
Fig.(\ref{fig:loopmatching}). It is clearly that the case 
in Eq.(\ref{OneloopHard}) is different from the one at 
tree level, $C_{k}^{(0)}\, =\, C_{\overline{k}}^{(0)}$. 
Parts of these QED corrections ($C_{k}^{\text {(a-d)}}$) 
are antisymmetric under the exchange of the collinear 
and anti-collinear sectors once hard fluctuations are 
decoupled.
The reason is that the exchange 
of the collinear and anti-collinear light quark is equivalent 
to performing the charge conjugation ({\it C}) just for 
final leptons as shown in the r.h.s of Fig.(\ref{fig:loopmatching}), 
that is, the matrix element, with one photon attached to one of 
leptons, will be transformed with {\it C} operator and 
simultaneously be matched onto $\mathrm{SCET_I}$ 
as
  \begin{align}\label{ChargeTrans1}
  \left\langle\gamma\, \left|Q_{i}, \mathcal{L}_{\mathrm{QED}}\, 
  \right| \, \ell^{+}\,  \ell^{-}\, \right\rangle 
  \, \stackrel{\mathrm{C} }{\longrightarrow}\, 
  -\, \left\langle 0\, \left|\, \mathcal{O}_{\bar i}\, 
  \right|\,  \ell^{+} \, \ell^{-}\, \right\rangle\, ,
  \end{align}
where operator $Q_{i}$ is changed into $Q_{\bar i}$ after 
{\it C} operator transition. Lagrangian $\mathcal{L}_{\mathrm{QED}}$ 
is {\it C} operator invariant, and the minus in r.h.s of above formula 
is from the action of {\it C} operator on one photon attached to the 
one of leptonic fields. It is possible to write Eq.(\ref{ChargeTrans1}) 
into a more general form to arbitrary loop order,
  \begin{align}\label{ChargeTransn}
  \left\langle \, m\, \gamma\, \left|\, Q_{i}, \mathcal{L}_{\mathrm{QED}}\,
   \right|\,  \ell^{+} \, \ell^{-}\, \right\rangle 
  \, \stackrel{\mathrm{C} }{\longrightarrow}\, 
  (-1)^m\,\left\langle \, 0\, \left|\, \mathcal{O}_{\bar i}\, 
  \right|\,  \ell^{+}\,  \ell^{-}\, \right\rangle\, ,
  \end{align}
where $m$ stands for the number of photon attached to 
lepton sector and determines whether the hard functions 
are symmetric or antisymmetric under the exchange of 
the collinear and anti-collinear light quark fields, that is, 
the relations, $C_{k}^{(m)}=(-1)^{m}\, C_{\bar{k}}^{(m)}, \, 
m=0,\, 1, \,2,\, ...\, ,$ are valid to all orders in 
$\alpha_{\mathrm{em}}$.

Based on these relations between $C_{k}^{(a-f)}$ and 
$C_{\bar k}^{(a-f)}$ in Eq.(\ref{OneloopHard}) and the 
one between $\mathcal{O}_9$ and $\mathcal{O}_{10}$ 
in Eq.(\ref{910relation}), the hard functions for $i=9$ at 
NLO in Eqs.(\ref{Hard9NLO2}) and (\ref{Hard9barNLO2}) 
can be simplified considerably as
  \begin{align}
  \begin{aligned}
  H^{(1)}_{9}\left(u, \mu_{b}\right)+
  H^{(1)}_{\bar 9}\left(u, \mu_{b}\right) 
  &=2 \mathcal{N}
  \left[\, C_{7,\, 9}^{(e-f)}\left(\mu_{b}\right) +
  C_{10}^{(a-d)}\left(\mu_{b}\right)\, \right]-
  \text{IR subtractions}\, 
  \end{aligned}
  \end{align}
which implied that $H^{(1)}_{9}$ equals to $H^{(1)}_{\bar 9}$ 
and the hard function can be written as
  \begin{align}\label{H91}
  H^{(1)}_{9/\bar 9}\left(u, \mu_{b}\right)\,&= \, \mathcal{N}\,
  \left[\, C_{7,\, 9}^{(e-f)}\left(\mu_{b}\right)\, +\, 
  C_{10}^{(a-d)}\left(\mu_{b}\right)\, \right]\, 
  -\, \text{IR subtractions}\, ,
  \end{align}
with the result as follows,
\begin{align}
  \begin{aligned}\label{H91result}
    H^{(1)}_{9/\bar 9}&\, = \, 
  (C_{9}^{\mathrm{eff}}-\frac{2 Q_\ell }{u} C_{7}^{\mathrm{eff}} )
  \left(\frac{\alpha_{\mathrm{em}}}{4 \pi} Q_b Q_q +
  \frac{\alpha_{\mathrm{s}}}{4 \pi}C_F\right)
  \left[-\ln ^{2} \frac{\tilde r}{\bar{u}}-2 \ln \frac{\tilde r}{\bar{u}}+
  \frac{1}{2} \ln ^{2} \tilde r+2  \mathrm{Li}_2 (-\frac{u}{\bar u})
  -4-\frac{\pi^{2}}{12}\right] \\ 
  &+(C_{9}^{\mathrm{eff}}-\frac{2 Q_\ell }{u}C_{7}^{\mathrm{eff}} )
  \frac{\alpha_{\mathrm{em}} }{ 4 \pi} Q_{\ell}^2\, 
   \left[-\ln ^{2} \frac{-u-i 0}{\tilde{r}}+3 \ln \frac{-u-i 0}{\tilde{r}}
   -8+\frac{\pi^{2}}{6}\right]\\
  &+ C_{9}^{\mathrm{eff}} \left( \frac{\alpha_{\mathrm{em}} }{ 4 \pi} 
  Q_{b} Q_{q} + \frac{\alpha_{\mathrm{s}}}{4 \pi} C_F\right)
  \left[\ln \tilde r-\frac{\bar{u}}{u} \ln \bar{u}\right] \\
  & + C_{10} \frac{\alpha_{\mathrm{em}}}{4 \pi} Q_{\ell} Q_{q} 
  \left[-\ln ^{2} \frac{u}{r}-\ln ^{2} \frac{-\bar{u}-i 0}{r}+\frac{2 \ln u}{\bar{u}}
  +\ln ^{2} r+3 \ln r+2 \operatorname{Li}_{2}\left(-\frac{\bar{u}}{u}\right)
  +10+\frac{\pi^{2}}{6}\right]\, ,
  \end{aligned}\end{align}
where $Q_\ell\, =\, -1,\,  Q_q\, =-\, 1/3$, 
$\tilde{r}\, =\, (\mu^{2}/m_{b}^{2})\, \mathrm{e}^{\gamma_{E}}$. 
The hard function $H^{(1)}_{9/\bar 9}$ has also contained 
QCD corrections which can be obtained with the replacement of 
$(\alpha_{\mathrm{em}}/ 4 \pi) \, Q_{b}\,  Q_{q}\,$ in QED contributions 
in above formula by $(\alpha_{\mathrm{s}} /4 \pi) \, C_F$.


\subsection{$\mathrm{HQET} \times \mathrm{bHLET}  $}
\label{HQETSCETI}

The first step of matching from $\mathrm{QCD} \times \mathrm{QED}$ 
onto $\mathrm{SCET_I}$ for $B_q\, \to \, \tau^+\, \tau^-$ described 
above is same as for $B_q \, \to \, \mu^+\,  \mu^-$. However, as for 
massive $\tau$ final states, 
$m_{\tau}^{2}\, \sim\, m_{b}\, \Lambda_{\mathrm{QCD}}$, the next 
matching from $\mathrm{SCET_I}$ will be different from the case 
in $\mu$ leptonic decays where the (anti-)hard-collinear leptons 
need to turn into (anti-)collinear ones by matching onto 
$\mathrm{SCET_{II}}$. The leptonic fields in $B_q\,  \to \, \tau^+\,  \tau^-$ 
decay should be converted to a soft-collinear field after integrating out the 
hard-collinear massive $\tau$ scale as done for massive $b$ quark fiels in the 
process $W \to B \gamma$ \cite{Beneke:2023nmj} in bHQET.
Consequently, in the following, we need to perform matching 
from $\mathrm{SCET_I}$ to $\mathrm{HQET} $ 
to turn the hard-collinear light antiquark to a soft one
to get a non-vanishing overlap $B$-meson state and
 simultaneously match onto $\mathrm{bHLET} $ to make the hard-collinear $\tau$ lepton
to be a soft-collinear one. 


\subsubsection[Operators]
 {\boldmath Operators}
\label{HQETJm}

The hard-collinear antiquark field of $\mathcal{O}_9$ in $\mathrm{SCET_I}$
can turn into a soft antiquark field through emission of a hard-collinear photon 
by the power-suppressed $\mathrm{SCET_I}$ Lagrangian,
  \begin{align}\label{lagrangiansub}
  \mathcal{L}_{\xi q}^{(1)}\, =\, 
  \bar{q}_{s}\left(x_{-}\right)\, \left[W_{\xi,\, hc}\,  W_{hc}\right]^{\dagger}(x) \, 
  i \, \slashed {D}_{hc \perp} \, \xi_{hc}(x)\, +\,  \mathrm{h.c.}\, 
  \end{align}
and analogously for anti-hard-collinear fields with the replacements of
$hc\,  \rightarrow\,  \overline{hc}, \, n_{+} \, \rightarrow \, n_{-}$ and 
$x_{-} \, \rightarrow \, x_{+}$. $W_{\xi,\,  hc}\, \text{and}\, W_{hc}$, 
connected with $ \xi_{hc}(x)$, are Wilson lines of hard-collinear 
photons and gluons, 
  \begin{align} 
  W_{\xi, hc}(x) & \, \equiv\, 
  \exp\,  \left[\, i\,  e\,  Q_{\xi}\,  \int_{-\infty}^{0}\,  d s\,
   n_{+} A_{hc}\left(x+s n_{+}\right)\, \right], \\ 
  W_{hc}(x) & \, \equiv \, 
  \mathcal{P}\,  \exp\,  \left[\, i\,  g_{s}\, \int_{-\infty}^{0} \, d s\, 
  n_{+} G_{hc}\left(x+s n_{+}\right)\, \right], 
\end{align}
respectively. Then the hard-collinear photon field, $A_{hc \perp}$ 
from $D_{hc \perp}$ in Eq.(\ref{lagrangiansub}), would be followed 
by the fusion, 
  \begin{align} \label{fusion}
  \bar{\ell}_{hc}\, +\, A_{\perp hc}\,  \stackrel{}{\rightarrow}\,  
  m_{\tau}\,  \bar{\ell}_{hc}\, ,
  \end{align}
through the leading power Lagrangian relevant to mass term, 
  \begin{align} 
  \mathcal{L}_{m}^{(0)}(y)&\, =\, m_{\tau}\,  \bar{\ell}_{C}\, 
  \left[\, i\,  \slashed D_{C \perp}\, , \, \frac{1}{i \, n_{+} \, D_{C}}\, \right]\,  
  \frac{\slashed n_{+}}{2}\,  \ell_{C}\, .
  \end{align}
 Therefore, we will match the time-ordered product of the  
$\mathrm{SCET_I}$ operators $\mathcal{O}_9(u)$ with 
$\mathcal{L}_{\xi q}^{(1)}(x)$ and $\mathcal{L}_{m}^{(0)}(y)$,
  \begin{align}\label{TimeorderSCETI1}
  \left\langle \, \ell^-(p_\ell)\,  \ell^+\left(p_{\bar{\ell}}\right)\, \left|\, 
  \int d^{4} x \, \int d^{4} y\, T\, \left\{\mathcal{O}_{9}(u), \, 
  \mathcal{L}_{\xi q}^{(1)}(x),\, \mathcal{L}_{m}^{(0)}(y) \right\}\, 
  \right|\,  b\left(p_{b}\right)\,  q\left(\ell_{q}\right)\, \right\rangle \, ,
  \end{align} 
to the corresponding matrix element of operator in 
$\text{$\mathrm{HQET} \times \mathrm{bHLET}$} $.

The systematic constructions of $\mathrm{HQET} \times \mathrm{bHLET}$ operators according to the analysis on 
a power-counting in $\lambda$, canonical dimension $d$, reparametrization 
symmetry of SCET, gauge symmetry and helicity conservation and so on, 
are similar to the ones performed in heavy-to-light meson form factors \cite{Beneke:2003pa} 
and the case of $B_q\, \to\, \mu^+\, \mu^-$ in Appendix B of \cite{Beneke:2019slt}. 
 At leading power, the $\mathrm{HQET} \times \mathrm{bHLET}$ operators used for the matching onto time-ordered product 
 in Eq.(\ref{TimeorderSCETI1}) can be defined in position space as
  \begin{eqnarray}\label{SECTIO9}
  \widetilde{\mathcal{J}}_{m \chi}^{A 1}(v)&\, =\, 
  \bar{q}_{s}\left(v  n_-\right) Y\left(v  n_-, 0\right)
  \frac{ \slashed n_-}{2}P_{L} h_{v}(0)
  [ Y_{+}^{\dagger} Y_{-} ](0)
  \left[ \bar{\ell}_{sc}(0)\, \left(4  P_{R}\right) 
  \ell_{\overline{sc}}(0)\right]\, ,
  \end{eqnarray}
 and 
  \begin{eqnarray}\label{SECTIO9bar}
  \widetilde{\mathcal{J}}_{m \bar \chi}^{A 1}(v)& =
  \bar{q}_{s}\left(v  n_+\right)  Y\left(v  n_+, 0\right)
  \frac{ \slashed n_+}{2} P_{L} h_{v}(0)
  [Y_{+}^{\dagger} Y_{-} ](0)
  \left[\bar{\ell}_{sc}(0) \left(4 P_{R}\right) 
  \ell_{\overline{sc}}(0) \right]\, ,
  \end{eqnarray}
for analogous operators generated from the matching relevant 
with $\mathcal{O}_{\bar 9}$, where the soft light antiquark field 
$\bar{q}_{s}\left(v  n_-\right) $ has been delocalized along the 
$n_-$ direction of light-cone through integrating out small 
component $n_- \ell_q$ of hard-collinear light quark. The 
roles of $n_{-}$ and $n_{+}$ are reversed for 
$\bar{q}_{s}\left(v  n_+\right)$ when the anti-hard-collinear 
mode is integrated out. $Y\left(v  n_+, 0\right)$ is the 
finite-distance Wilson line defined as
  \begin{eqnarray}
  Y(x, y)\, =\, \exp\,  
  \left[\, i \, e\,  Q_{q} \, \int_{y}^{x} \, d z_{\mu}\, A_{s}^{\mu}(z)\, \right] \, 
  \mathcal{P}\,  \exp\, 
   \left[\, i\,  g_{s}\,  \int_{y}^{x}\,  d z_{\mu}\, G_{s}^{\mu}(z)\, \right]\, ,
  \end{eqnarray}
to connect non-local field $\bar{q}_{s}\left(v  n_+\right)$ to $ h_{v}(0)$
to maintain QCD and QED gauge invariance of non-local operator. 
Here $\mathcal{P}$ is the path-ordering operator. $A_{s}^{\mu}(z)$ 
and $G_{s}^{\mu}\, =\, G_{s}^{\mu A}\,  T^{A}$ are the soft photon 
and gluon fields, respectively. The product of Wilson lines 
$\left[\, Y_{+}^{\dagger} Y_{-}\, \right](0)\,  \equiv\,  
Y_{+}^{\dagger}(0)\,  Y_{-}(0)$ appears after decoupling of soft 
photons from the hard-collinear and anti-hard-collinear leptons
in $\mathrm{SCET}_{\mathrm{I}}$, respectively, with their small 
component $n_- p_\ell$ and $n_+ p_{\bar \ell}$ scaling as 
$\lambda^2$ in the same order as soft photons. The soft 
electromagnetic Wilson lines are defined as
  \begin{eqnarray}
  Y_{\pm}(x)\, =\,  \exp\,  \left[\, -i \, e \, Q_{\ell}\,  \int_{0}^{\infty}\,  d s\, 
  n_{\mp} A_{s}\left(x+s n_{\mp}\right)\, \right]\, .
  \end{eqnarray}
For $\tau$ final states here, there is no so-called $B1$-type operator
as appeared in the case of $\mu$ leptons attributing to collinear 
contribution. It is only one way as Eq.(\ref{fusion}) to built 
$A1$-type operator in $\text{$\mathrm{HQET} \times \mathrm{bHLET}$}$.
  
We define the Fourier transforms of $\widetilde{\mathcal{J}}_{i}^{A 1}(v)$
in order to do matching from $\mathrm{SCET_I}$ to 
$\text{$\mathrm{HQET} \times \mathrm{bHLET}$}$ in momentum space as
  \begin{eqnarray}
  \mathcal{J}_{i}^{A 1}(\omega)\, =\, \int\,  
  \frac{d v}{2 \pi} \, e^{i\,  \omega\,  v}\, \widetilde{\mathcal{J}}_{i}^{A 1}(v)\, ,
  \end{eqnarray}
where $\omega$ corresponds to the soft momentum of the light quark 
along the $n_+$ and $n_-$ direction for $i=m \chi\, \text{and}\, m \bar\chi $, 
respectively. 
 
 \subsubsection{Matching  from $\mathrm{SCET_I}$ to 
                          $\text{HQET}\times \text{bHLET}$}
\label{sec3.3.2} 
\begin{figure}[h]
  \begin{center}
  \vspace{0 cm}\hskip0cm
  \includegraphics[width=0.5\textwidth]{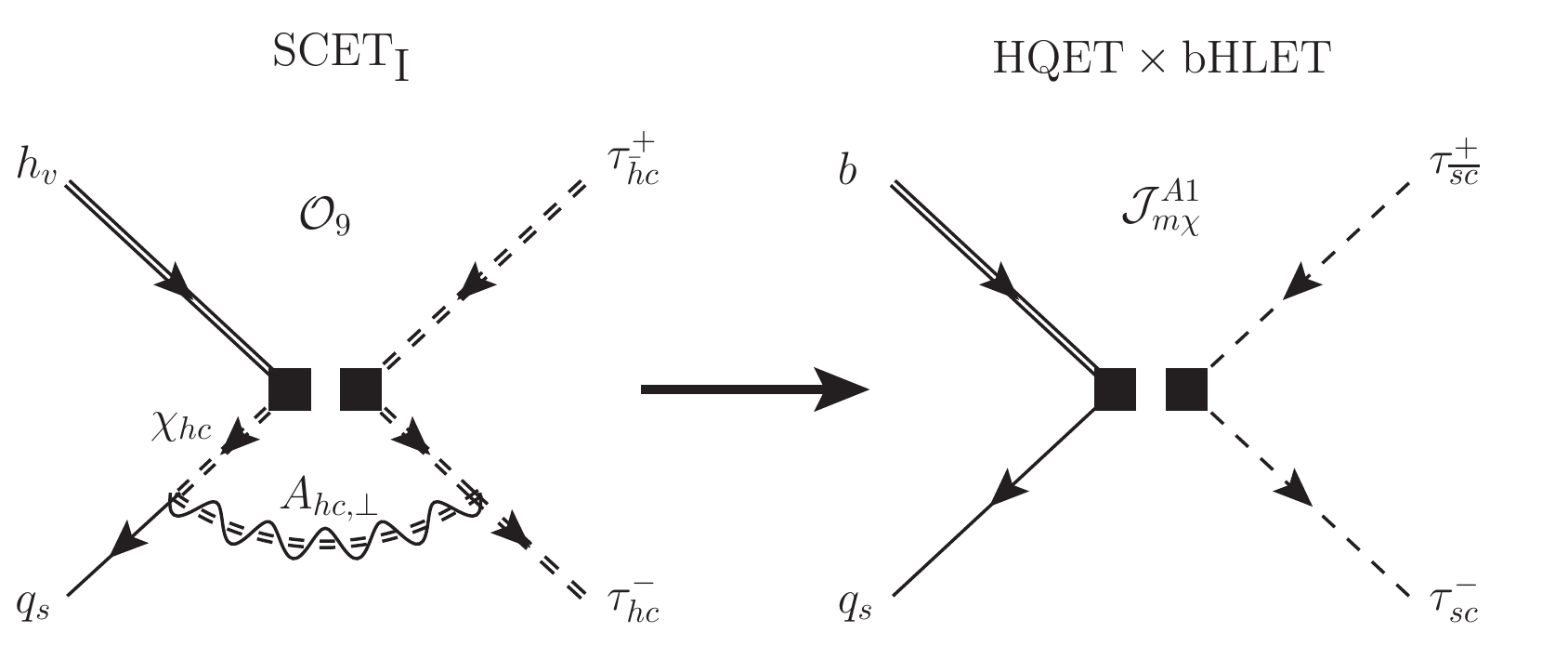}
  \end{center}
  \caption{The figure shows the tree-level matching of 
  $\mathrm{SCET_I}$ onto $\text{$\mathrm{HQET} \times \mathrm{bHLET}$}$, 
  where $\mathrm{SCET_I}$ operator $\mathcal{O}_{9}$ connected 
  with two currents forms a time-ordered product to match onto 
  $A1$-type operator $\mathcal{J}_{m \chi}^{A 1}$. The 
  double-dashed lines accompanied by a wavy line depict 
  the hard-collinear photon field $A_{hc,\, \perp}$ in 
  $\mathrm{SCET_I}$. The single-solid line and single-dotted line in 
  $\mathcal{J}_{m \chi}^{A 1}$ represent the soft spectator quark 
  field $q_s$ and soft-collinear $\tau$ fields.}
  \label{fig:treematching}
  \end{figure}

We perform the matching of operator $\mathcal{O}_{9}$ in 
$\mathrm{SCET_I}$ onto $\mathcal{J}_{m \chi}^{A 1}$ in 
$\text{$\mathrm{HQET} \times \mathrm{bHLET}$} $ at hard-collinear scale $\mu_{hc}$,
  \begin{eqnarray}\label{matchingequationA}
  \mathcal{O}_{9}(u)&\,  \rightarrow \,  \int \, d \omega\,  
  J_{m}(u ; \omega )\,  \mathcal{J}_{m \chi}^{A 1}(\omega)\, .
  \end{eqnarray} 
 As mentioned in Sec.\ref{HQETJm}, the operator in l.h.s of 
 Eq.(\ref{matchingequationA}) should be connected with two 
 currents $\mathcal{L}_{\xi q}^{(1)}(x)\, \text{and}\, 
 \mathcal{L}_{m}^{(0)}(y)$ to produce a non-vanishing overlap 
 $B$-meson state and bHLET modes. The tree-level matching relation is depicted
 in Fig.(\ref{fig:treematching}).
Firstly, we need to calculate the matrix element of the time-ordered product of the  
$\mathrm{SCET_I}$ operators $\mathcal{O}_9(u)$ with 
$\mathcal{L}_{\xi q}^{(1)}(x)$ and $\mathcal{L}_{m}^{(0)}(y)$,
  \begin{align}\label{TimeorderSCETI}
  \langle \, \ell^-(p_{\ell^-})\,  \ell^+\left(p_{\ell^+}\right)\, |\, \int d^{4} x \, 
  \int d^{4} y\, T\, \{\mathcal{O}_{9}(u), \, \mathcal{L}_{\xi q}^{(1)}(x),\,  
  \mathcal{L}_{m}^{(0)}(y) \} \, | b\left(p_{b}\right) \, q\left(\ell_{q}\right)\, \rangle \, ,
  \end{align}
and the result is 
  \begin{align}\label{TimeorderSCETIResult}
   \frac{\alpha_{\mathrm{em}}}{4\pi}\,  Q_{\ell} \, Q_{s}\,  
   m_\ell\,  \frac{\bar u}{\omega }\, \ln\,  [\, 1\, +\, \frac{u}{\bar u}\, 
   \frac{\omega\, n_+ p_{\ell^-}}{m_\ell^2} \, ]\, 
   \theta(u)\,  \theta(\bar u) \, \, 
   [\, \bar {q}_s\,  \gamma^\perp_\nu\, \gamma^\perp_\mu\,  
   \frac{\slashed{n}_-}{2}\, P_L\,  h_v\, ] \,
   [\, \bar {\ell}_{sc}\, \, \gamma_\perp^\nu\,  \gamma_\perp^\mu \, 
   \ell_{\overline {sc}} \, ]\, .
  \end{align}
We can define a matrix element of an operator denoted by 
$\mathcal{J}_{9}^{A1}$ as
  \begin{align}\label{J9A1}
  \langle\mathcal{J}_{9}^{A1}\rangle \,  \equiv \, 
  \left[\, \bar {q}_s (l_q)\,  \gamma^\perp_\nu\, \gamma^\perp_\mu\, 
  (\slashed{n}_-/2)\, P_L\,  h_v(p_b)\, \right] \, 
  \left[\, \bar {\ell}_{sc} (p_{\ell^-})\, \gamma_\perp^\nu\,  \gamma_\perp^\mu \, 
  \ell_{\overline {sc}} (p_{\ell^+}) \, \right]\, .
 \end{align}
$\mathcal{J}_{9}^{A1}$ is equal to $\mathcal{J}_{m \chi}^{A 1}$ in 
dimension $d=4$,
  \begin{align}
  \mathcal{J}_{9}^{A1} \, \stackrel{\mathrm{d=4}}{\longrightarrow} \, 
  \mathcal{J}_{m \chi}^{A 1}\, ,
  \end{align}
so that we can define an evanescent operator as
  \begin{align}
  \mathcal{J}_{E}^{A1}\, =\, 
  \mathcal{J}_{9}^{A1}\, -\, \mathcal{J}_{m \chi}^{A 1}\, .
  \end{align}
Hard-collinear function $J_m(u,\omega)$ can be extracted from 
the matching equation at tree level,
  \begin{align}
  C_{9m}^{(0)} \, \langle\mathcal{J}_{9}^{A1} (u)\rangle^{(0)}\,
   =\, \int\,  d \omega \, \left[\, J_{m}^{(0)}(u ; \omega)\, 
   \langle\mathcal{J}_{m \chi}^{A 1}(\omega)\rangle^{(0)}\, +\, 
   J_{E}^{(0)} (u ; \omega)\, \langle \mathcal{J}_{E}^{A1}(u) \rangle^{(0)}\, \right]\, ,
  \end{align}
 where the coefficient $C_{9m}^{(0)} \,$ can be read from the result 
 of Eq.(\ref{TimeorderSCETIResult}),
  \begin{align}
  C_{9m}^{(0)} \,= \, \frac{\alpha_{\mathrm{em}}}{4\pi}\, Q_{\ell} \, Q_{s}\,  
 m_\ell\,  \frac{\bar u}{\omega }\, \ln\,  \left(\, 1\, +\, \frac{u}{\bar u}\,  
  \frac{\omega\, n_+ p_{\ell^-}}{m_\ell^2}\,  \right)\, \theta(u)\,  \theta(\bar u) \, .
  \end{align}
At tree level, the hard-collinear function $J_m(u,\omega)$ is 
just $C_{9m}^{(0)}$, 
  \begin{align}\label{TimeorderSCETIresult3}
  J_{m}^{(0)}(u ; \omega; \mu= \mu_{hc})&\, =  \, 
  \frac{\alpha_{\mathrm{em}}}{4\pi}\, Q_{\ell} \, Q_{s}\,  
m_\ell\,  \frac{\bar u}{\omega }\, \ln \, \left(\, 1\, +\, \frac{u}{\bar u}\,  
  \frac{\omega\, n_+ p_{\ell^-}}{m_\ell^2} \, \right)\, 
  \theta(u) \, \theta(\bar u) \, ,
  \end{align}
at hard-collinear scale $\mu_{hc}$, and the result for 
$J_{\overline{m}}^{(0)}$ can be obtained by the replacement 
of $n_{+} p_{\ell^-} \, \rightarrow\,  n_{-} p_{\ell^+}$.

\subsection{Matrix elements of operators 
                   $\widetilde{\mathcal{J}}_{m \chi, \bar \chi}^{A 1}$ in 
                   $\text{$\mathrm{HQET} \times \mathrm{bHLET}$} $}
\label{softmatrix}

The operators $\widetilde{\mathcal{J}}_{m \chi, \bar \chi}^{A 1}$ in
Eqs.(\ref{SECTIO9}) and (\ref{SECTIO9bar}) are composed of soft, 
soft-collinear sector and anti-soft-collinear sector, and these fields 
do not interact with one another, which implies that the matrix elements 
of the operators $\widetilde{\mathcal{J}}_{m \chi, \bar \chi}^{A 1}$ can 
be further factorized accordingly into matrix elements of the separate 
factors in the respective soft, soft-collinear and anti-soft-collinear 
Hilbert space,
  \begin{align}\label{Softmatrix}
  \langle\,  \ell^+\, \ell^-\, |\, \widetilde{\mathcal{J}}_{m \chi}^{A 1}\,
   |\, \bar B_q \, \rangle\, =\, 
  \langle\, 0\, |\, \widehat{\mathcal{J}}_{s} \, |\, \bar B_q\, \rangle\,
  \langle\, \ell^-\, |\, \widehat{\mathcal{J}}_{sc} \, |\, 0\,  \rangle\, 
 \langle\, \ell^+\, |\, \widehat{\mathcal{J}}_{\overline{sc}}\, |\, 0\, \rangle\, .
 \end{align}
Then the soft, (anti-)soft-collinear sectors are defined as
  \begin{align}
  \widehat{\mathcal{J}}_{s}&\, =\, \bar{q}_{s}\left(v n_{-}\right)\,  
  Y\left(v n_{-}, 0\right)\,  \frac{\not \! n_-}{2} \, 
  P_{L}\,  h_{v}(0)\, [Y_{+}^{\dagger}\, Y_{-} ](0)\, ,\\
  \widehat{\mathcal{J}}_{sc}&\, =\, \bar{\ell}_{sc}(0)\, 
  \left( 4 P_{R}\right)\, , \, \quad
  \widehat{\mathcal{J}}_{\overline{sc}}\, = \, \ell_{\overline{sc}}(0)\, .
  \end{align}
However, when considering the renormalization of each sector 
separately, one find IR-divergence can not be cancelled only in 
soft sector, and the remaining divergence need to be cancelled 
by the one from (anti-)soft-collinear sectors (more details can be 
found in Section 4.2 of Ref.~\cite{Beneke:2019slt}). This appears 
to be in conflict with the factorization of the soft and 
(anti-)soft-collinear sectors. It is so-called factorization anomaly, 
which will lead to a rearrangement for soft operator. In order to 
subtract the remaining IR-divergence in soft operator, one can 
redefine and renormalize soft function as 
  \begin{align}
  \widetilde{\mathcal{J}}_{s}(v) \, \equiv \, \frac{\widehat{\mathcal{J}}_{s}(v)}
   {\, \langle\,  0 \, | \, [\, Y_{+}^{\dagger} \, Y_{-} \, ](0)\, | \, 0 \, \rangle}\, ,
  \end{align}
where $\langle\,  0\,  |\,  [\, Y_{+}^{\dagger} \, Y_{-} \, ]\, (0)\, | \, 0 \, \rangle$ 
in the denominator is just the overlap term between soft and (anti-)soft-collinear 
regions. It can be divided into two separate factors $R_+$ and $R_-$ by 
$\, \langle\, 0\, |\, [\, Y_{+}^{\dagger} \, Y_{-}\, ]\, (0)\, |\,  0\, \rangle \, \equiv \, 
R_{+} \, R_{-}$, where $R_+$ and $R_-$ can be chosen as a symmetric form 
upon exchanging $n_+\leftrightarrow n_-$. Then soft-collinear sector and 
anti-soft-collinear sector are redefined correspondingly as 
  \begin{align}\label{hcoperator}
  \widetilde{\mathcal{J}}_{sc}&\, =\, 
  R_{+}\, \overline{\ell}_{sc}(0)\, \left(\, 4 \, P_{R}\, \right)\, , \, 
  \quad
  \widetilde{\mathcal{J}}_{\overline{sc}}\, =\, R_{-}\, \ell_{\overline{sc}}(0)\, .
  \end{align}
  
The hadronic matrix element of the soft operator 
$\, \langle \, 0\, |\, \widetilde{\mathcal{J}}_{s} \, |\, B_q \, \rangle$ 
is related to  the $B_q$ meson decay constant and the leading-twist 
$B_q$ meson LCDA \cite{Grozin:1996pq,Beneke:2000wa}. However, 
it would not coincide with the universal $B_q$ meson LCDA but 
depend on the final-state particles of the specific process due to 
appearance of additional soft QED Wilson lines 
$ Y_{+}^{\dagger}\, Y_{-} $. Therefore, we define the 
$\langle \, 0\, |\, \widetilde{\mathcal{J}}_{s}\, |\, B_q \, \rangle$ as 
a generalized and process-dependent $B_q$ meson LCDA 
$\Phi_{+}(\omega)$, 
  \begin{align}\label{BLCDA}
  \begin{aligned}
   (-4)\, \langle \, 0 \, |\, \widetilde{\mathcal{J}}_{s}(v) \, |\,  \bar{B}_{q}(p)\, \rangle 
    &\, =\, \frac{\, \langle\,  0\, |\, \bar{q}_{s}\left(v n_{-}\right)\,  Y\left(v n_{-}, 0\right)
     \, \not\! n_- \, \gamma_{5} \, h_{v}(0)\,  Y_{+}^{\dagger}(0) \, Y_{-}(0)\, | \,
     \bar{B}_{q}(p)\, \rangle}{\, \langle \, 0\,  |\, [\, Y_{+}^{\dagger}\,  Y_{-}\, ]\, (0)\, 
     | \, 0\, \rangle} \\ 
    &\, \equiv\,  i \, m_{B_{q}}\,  \int_{0}^{\infty}\,  d \omega\, e^{-i \, \omega \, v}\, 
     \mathcal{F}_{B_{q}}\,  \Phi_{+}(\omega)\, .
    \end{aligned}
    \end{align}
The analogous definition holds for the anti-collinear case by 
interchanging $n_+\, \leftrightarrow \, n_-$ but with the same 
$B_q$ meson LCDA $\Phi_{+}(\omega)$. $\mathcal{F}_{B_{q}}$ 
is the generalized process-dependent $B_q$ meson decay 
constant, which can be defined through the local matrix element,
  \begin{align}\label{BDC}
  \frac{\, \langle \, 0\,  |\, \bar{q}_{s}(0)\,  \gamma^{\mu}\,  \gamma_{5} \, 
  h_{v}(0) \, Y_{+}^{\dagger}(0) \, Y_{-}(0) \, |\, \bar{B}_{q}(p)\, \rangle} 
  {\, \langle\,  0 \, |\, [\, Y_{+}^{\dagger} \, Y_{-}]\, (0)\, | \, 0\, \rangle}
  \, \equiv\,  i \, \mathcal{F}_{B_{q}}\,  m_{B_{q}}\,  v^{\mu}\, .
  \end{align}
    
Although the specific process-dependent $B_q$ meson LCDA and 
decay constant in Eqs.(\ref{BLCDA}) and (\ref{BDC}) are complicated 
in the presence of QED effects, we can expand them perturbatively 
in terms of $\alpha_{\mathrm{em}}$ at the soft scale 
$\mu_s\, \sim\, \Lambda_{\mathrm{QCD}}$,
  \begin{align}\label{BmesonLCDA}
  \mathcal{F}_{B_{q}}\left(\mu_{s}\right) &
  \, =\, \sum_{n=0}^{\infty}\, \left(\, \frac{\alpha_{\mathrm{em}}\left(\mu_{s}\right)}
  {4\,  \pi}\right)^{n} \, F_{B_{q}}^{(n)}\left(\mu_{s}\right)\, , \\
  \mathcal{F}_{B_{q}}\left(\mu_{s}\right) \, \Phi_{+}\left(\omega ; \mu_{s}\right) 
  &\, =\, \sum_{n=0}^{\infty}\, \, \sum_{k=0}^{\infty}\, \left(\, \frac{\alpha_{\mathrm{em}}
  \left(\mu_{s}\right)}{4\,  \pi}\right)^{n+k} \, F_{B_{q}}^{(n)}\left(\mu_{s}\right) \, 
  \phi_{+}^{(k)}\left(\omega ; \mu_{s}\right) \, ,
  \end{align}
where the leading terms, $n=0$ and $k=0$, are just the standard $B_q$ meson 
decay constant $F_{B_{q}}(\mu_s)$ and 
$\operatorname{LCDA}, \phi_{+}(\omega ; \mu_s)$, defined in the 
absence of QED correction. Higher-order terms starting from $n=1$ or $k=1$
in the expansion with QCD and QED correction simultaneously are 
non-universal, non-local HQET matrix elements that have to be 
evaluated nonperturbatively. Fortunately, only the universal objects 
$F_{B_{q}}(\mu_s)$ and $\phi_{+}(\omega ; \mu_s)$ need to be 
known at the leading and next-to-leading logarithmic (NLL) accuracy.

Next the matrix element of (anti-)soft-collinear in Eq.(\ref{hcoperator}) 
can be defined after renormalization by 
  \begin{align}\label{hcMatrix}
  \left\langle \ell^{-}\left(p_{\ell^-}\right)
  \left| R_{+}  \bar{\ell}_{sc}(0) \right| 0\right\rangle
   =Z_{\ell} \bar{u}_{sc}\left(p_{\ell^-}\right), 
  \quad
  \left\langle \ell^{+}\left(p_{\ell^+}\right)
  \left| R_{-}  \ell_{\overline{sc}}(0) \right| 0 \right\rangle
  = Z_{\bar{\ell}} v_{\overline{sc}}\left(p_{\ell^+}\right)\, ,
  \end{align}
where
  \begin{align}
  Z_{\ell}\, =\, Z_{\bar{\ell}}\, =\, 
  1\, +\, \mathcal{O}\left(\alpha_{\mathrm{em}}\right)\, .
  \end{align}

Collecting the soft and (anti-)soft-collinear sectors in 
Eqs.(\ref{BLCDA}) and (\ref{hcMatrix}), respectively, 
we can now derive the factorized expression for the 
matrix element of Eq.(\ref{Softmatrix}) in the Fourier 
transformed form,
  \begin{align}\label{SoftMatrix}
  \left\langle\, \ell^{+}\left(p_{\ell^+}\right) \, \ell^{-}\left(p_{\ell^-}\right)\, 
  \left|\, \mathcal{J}_{m \chi}^{A 1}(\omega)\, \right|\,  
  \bar{B}_{q}(p)\, \right\rangle
  \, =\, T_{+}\, m_{B_{q}}\, \mathcal{F}_{B_{q}}\, \Phi_{+}(\omega)\, ,
  \end{align}
where
  \begin{align}\label{Tplus}
  T_{+}(\mu) \, \equiv\, 
  (-i)\, m_{\ell}(\mu)\, Z_{\ell}(\mu)\,  Z_{\bar{\ell}}(\mu)\, 
  \left[\, \bar{u}_{sc}\left(p_{\ell^-}\right) \, P_{R} \, 
  v_{\overline{sc}}\left(p_{\ell^+}\right)\, \right]\, .
  \end{align}
The same result holds for the anti-soft-collinear operators 
$\mathcal{J}_{m \bar\chi}^{A 1}$ owning to the same 
soft matrix element definition in Eq.(\ref{BLCDA}).

\section{Resummed amplitude of $B_{q}\, \rightarrow\, \tau^{+}\, \tau^{-}$}
\label{RGinvolution}

\subsection{Factorization of the amplitude}

After factorizing $B_{q} \, \rightarrow \, \tau^{+} \, \tau^{-}$ 
decay into three parts: hard function, (anti-)hard-collinear 
function and soft function, in Sec.\ref{Sec3}, we can thus 
write the complete expression of its amplitude by adding 
the hard function and hard-collinear matching coefficients 
to Eq.(\ref{SoftMatrix}) as
  \begin{align}\label{Factorized amp}
  i \, \mathcal{A}_{9}&=T_{+} \int_{0}^{1} d u 
  \left[H_{9}^{(0)}(u, \mu)+ H_{\bar 9}^{(0)}(u, \mu)+2 H^{(1)}_{9/\bar 9}(u, \mu)\right]
  \int_{0}^{\infty} d \omega J_{m}(u ; \omega, \mu)
  m_{B_{q}} \mathcal{F}_{B_{q}}(\mu) \Phi_{+}(\omega, \mu),\nonumber\\
  &=\, T_{+}\,  \int_{0}^{1} d u \, 
 2 H_{9}(u, \mu)\, 
  \int_{0}^{\infty}\,  d \omega\,  J_{m}(u ; \omega, \mu)\, 
  m_{B_{q}} \, \mathcal{F}_{B_{q}}(\mu) \, \Phi_{+}(\omega, \mu)\, ,
  \end{align}
where $H^{(0)}_9 ,\, H^{(0)}_{\bar 9} ,\, H^{(1)}_{9/\bar 9}$ are shown in Eqs.(\ref{H90}),
(\ref{H90bar}) and (\ref{H91result}), respectively. The summation of them can be simplified as
\begin{align}\label{hardH9}
2\, H_9\left(u, \mu_{b}\right)\,&=  2\mathcal{N}\, 
  \left[\, C_{9}^{\mathrm{eff}}\left(u, \mu_{b}\right)\, 
  -\, \frac{2 \, Q_{\ell}}{u} \, C_{7}^{\mathrm{eff}}\left(u, \mu_{b}\right)\, \right]\, +\, 2 H^{(1)}_{9/\bar 9}\, ,
  \end{align}
 where the contribution $C^{(0)}_{10}$ from operator $Q_{10}$ in Eqs.(\ref{H90}) 
and (\ref{H90bar}) has cancelled.
Scale $\mu$ in the amplitude can be chosen 
to be an arbitrary but the same scale, which will inevitably lead 
to the presence of large logarithms as functions of multi-scales. 
In order to avoid the large logarithm, we evaluated each factor 
in Eq.(\ref{Factorized amp}) at their intrinsic scales, that is, 
hard function $H_9 (u, \mu)$ at hard 
scale $\mu_h\, \sim\, m_b$, hard-collinear function 
$J_{m}(u ; \omega, \mu)$ at hard-collinear scale 
$\mu_{hc}\, \sim\, \sqrt {m_b\,  \Lambda_{\mathrm{QCD}}}$, 
and soft $B$-LCDA had also been calculated by some 
nonperturbative approaches, such as light-cone sum rules, 
Lattice QCD, or extracted from experimental data directly.
Next, it is necessary to make each factor in 
Eq.(\ref{Factorized amp}) run to a common scale 
$\mu$ by renormalization group equations (RGE).

\subsection{Resummed amplitude}
 
We will apply the solutions to RGE to convert Eq.(\ref{Factorized amp}) 
into the one where hard function, hard-collinear function and soft function 
would have run to a common scale, meanwhile, the large logarithms 
would have been resummed. The explicit result of resummation will be 
given in the leading logarithms (LL) approximation and we shall choose 
the common scale in Eq.(\ref{Factorized amp}) to be the hard-collinear 
scale $\mu_{hc}\, \sim\, \sqrt {m_b\, \Lambda_{\mathrm{QCD}}}$.

\subsubsection{The evolution of Hard function}
 
The RGE in $\mathrm{SCET_I}$ governs the evolution of the 
hard function $H_{9} (u, \mu)$ from the hard scale 
$\mu_b$ down to the hard-collinear scale $\mu_{hc}$. 
It requires that we need to know the anomalous dimensions 
of operator $\mathcal {O}_{9}$. The evaluation of the 
anomalous dimensions of $\mathcal {O}_{9}$ is similar 
with the one of N-jet operators
\cite{Beneke:2017ztn, Beneke:2018rbh}, and it has also been 
performed in the process of $B_q \, \to\,  \mu^+\,  \mu^-$. 
The evolution of the hard function of $H_{9}$ can be expressed as
  \begin{eqnarray}\label{Hardevolu}
  H_{9}(u, \mu)\, = \, U_{h}\left(\mu_{b}, \mu\right) 
   \, H_{9}\left(u, \mu_{b}\right)\, .
   \end{eqnarray}
In the LL approximation, that is, only cusp anomalous dimensions 
are kept, the evolution function is
  \begin{align}
   U_{h}\left(\mu_{b}, \mu \right) &\, =\, \exp\,  
   \left[\, \int_{\mu_{b}}^{\mu} \, \frac{d \mu^{\prime}}{\mu^{\prime}} \, 
   \Gamma^{\mathrm{I}}_{\mathrm{cusp}}\left(\mu^{\prime}\right)\, 
   \ln\,  \frac{m_{B_{q}}}{\mu^{\prime}}\, \right]\, ,
   \end{align}
with cusp anomalous dimensions
  \begin{align}
  \Gamma_{\mathrm{cusp}}^{\mathrm{I}}
  \left(\alpha_{s}, \alpha_{\mathrm{em}}\right)
  &\, =\,  \frac{\alpha_{s}}{\pi}\,  C_{F}\, +\, 
  \frac{\alpha_{\mathrm{em}}}{\pi}\,  \left[\, Q^2_{q}\, +\, 
  +\, 2 \, Q_{q}\, Q_{\ell}\, +\, 2 \, Q_{\ell}^{2}\, \right]\, .
  \end{align}

At the next-to-leading logarithmic (NLL) accuracy, one 
would also consider one-loop anomalous dimension
  \begin{align}
  \Gamma_{i}(x, y)\, =\, \frac{\alpha_{s} C_{F}}{4 \pi}\, 
  [\, 4\,  \ln\,  (1-x)\, -\, 5\, ] \, \delta(x-y)\, +\, 
  \frac{\alpha_{\mathrm{em}}}{4 \pi} \, \gamma_{i}(x, y), \quad(i=9)\, ,
  \end{align}
and two-loop cusp anomalous dimension in the evolution 
equation,
  \begin{align}
  \frac{d H_{i}(u, \mu)}{d \ln \mu}\, =\, \Gamma_{\text {cusp }}^{\mathrm{I}}\, 
  \left(\, \ln\,  \frac{m_{B_{q}}}{\mu}\, -\, \frac{i \, \pi}{2}\, \right) \,
  H_{i}(u, \mu)\, +\, \int d u^{\prime}\,  \Gamma_{i}\left(u^{\prime}, u\right) \, 
  H_{i}\left(u^{\prime}, \mu\right)\, .
  \end{align}
The one-loop anomalous dimension $\gamma_{i}(x, y)$ is provided 
here for completeness,
  \begin{align}
  \gamma_{i}(x, y)\, =\, 
  &\delta(x-y)\, \left[\, Q_{\ell}^{2}\, (4\,  \ln x\, -\, 6)\, +\, 
  Q_{\ell}\,  Q_{q}\,  4\,  \ln x\, \bar{x}\, +\, 
  Q_{q}^{2}\, (4\,  \ln\,  \bar{x}\, -\, 5\, )\, \right]\, +\, 
  \nonumber\\
  &4 Q_{\ell} \, Q_{q}
   \left[\frac{\bar{x}}{\bar{y}} \left(\left[\frac{\theta(x-y) }{x-y}\right]_{+}+\theta(x-y) \right)
  + \frac{x}{y}\left(\left[\frac{\theta(y-x)}{y-x}\right]_{+}+\theta(y-x)\right)\right]\, ,
  \end{align}
where $[...]_+$ is the plus function. 

\subsubsection{The evolution of soft function}
 
Next we have to evolve the soft functions up from 
$\mu_s\,  \sim \, \Lambda_{\mathrm{QCD}}$ to $\mu_{hc}$. 
With the addition of the solution of the soft evolution equation, 
the soft matrix element at arbitrary scale in the LL approximation 
is
  \begin{align}\label{softQED}
  \mathcal{F}_{B_{q}}(\mu)\,  \Phi_{+}(\omega ; \mu)\, 
  =\, U_{s}\left(\mu, \mu_{s} ; \omega\right)\,  
  \mathcal{F}_{B_{q}}\left(\mu_{s}\right)\, 
  \Phi_{+}\left(\omega ; \mu_{s}\right)\, .
  \end{align}
It is useful to divide the evolution function into the QED 
and QCD parts,
  \begin{align}
  U_{s}\left(\mu, \mu_{s} ; \omega, \omega^{\prime}\right)\, =\, 
  U_{s}^{\mathrm{QCD}}\left(\mu, \mu_{s} ; \omega, \omega^{\prime}\right) \, 
  U_{s}^{\mathrm{QED}}\left(\mu, \mu_{s} ; \omega, \omega^{\prime}\right)\, ,
  \end{align}
where $U_{s}^{\mathrm{QCD}}\left(\mu, \mu_{s} ; \omega, \omega^{\prime}\right)\,$
is the evolution factor for the standard $B$-meson LCDA in 
the absence of QED. Then Eq.(\ref{softQED}) can be written 
as
  \begin{align}\label{evolutionSF}
  \begin{aligned} \mathcal{F}_{B_{q}} \Phi_{+}(\omega) &\, =\,  
  U_{s}^{\mathrm{QED}}\left(\mu, \mu_{s} ; \omega, \omega^{\prime}\right) \, 
  U_{s}^{\mathrm{QCD}}\left(\mu, \mu_{s} ; \omega, \omega^{\prime}\right)\,
  \mathcal{F}_{B_{q}}\left(\mu_{s}\right)\,  \Phi_{+}\left(\omega ; \mu_{s}\right)\, \\ 
  &\, \rightarrow \, 
  U_{s}^{\mathrm{QED}}\left(\mu, \mu_{s} ; \omega, \omega^{\prime}\right) \, 
  U_{s}^{\mathrm{QCD}}\left(\mu, \mu_{s} ; \omega, \omega^{\prime}\right)\,
  F_{B_{q}}\left(\mu_{s}\right)\,  \phi_{+}\left(\omega ; \mu_{s}\right)\,  \\ 
  &\, = \, 
  U_{s}^{\mathrm{QED}}\left(\mu_{h c}, \mu_{s} ; \omega\right) \, 
  F_{B_{q}}\left(\mu_{h c}\right)\,  \phi_{+}\left(\omega ; \mu_{h c}\right)\, ,
  \end{aligned}
  \end{align}
where the second arrow represents that the leading order of 
Eq.(\ref{BmesonLCDA}), e.g. the standard $B$-meson LCDA 
$\phi_{+}$ and HQET decay constant $F_{B_{q}}$, has only 
been kept in LL accuracy. 

The $B$-meson LCDA $\Phi_{+}(\omega ; \mu)$ fulfils the RGE 
as \cite{Beneke:2022msp}
  \begin{align}
  \frac{d}{d \ln \mu} \Phi_{+}(\omega ; \mu)\, =\, 
   -\int_{-\infty}^{\infty}\, d \omega^{\prime}\,  \Gamma_{s}  
   \left(\omega, \omega^{\prime} ; \mu\right) \, 
   \Phi_{+} \left(\omega^{\prime} ; \mu\right)
   \end{align}
 with the anomalous dimension given by
  \begin{align}\label{anomalousdim}
  \Gamma_{s}\left(\omega, \omega^{\prime} ; \mu\right)\, =\, 
  -\int_{-\infty}^{\infty} d \hat{\omega}\, \frac{d Z_{\otimes}(\omega, \hat{\omega} ; \mu)}
  {d \ln \mu} \, Z_{\otimes}^{-1}\left(\hat{\omega}, \omega^{\prime} ; \mu\right)
  \, +\, \delta\left(\omega-\omega^{\prime}\right)\, \frac{d \mathcal{F}_{B_{q}}(\mu)}{d \ln \mu}\, .
  \end{align}
The results for the anomalous dimension of Eq.(\ref{anomalousdim}) 
can be found in Eq.(2.23) of \cite{Beneke:2022msp}. Here for the LL accuracy, 
we only keep the cusp part of the anomalous dimension and the result 
of QED evolution functions $U_{s }^{\mathrm{QED}}\left(\mu_{h c}, 
\mu_{s} ; \omega\right)$ is (see also, ref. \cite{Beneke:2019slt})
  \begin{align}\label{Softevolu}
  U_{s }^{\mathrm{QED}}\left(\mu_{h c}, \mu_{s} ; \omega\right)
  &=\exp \left[\frac{4 \pi}{\alpha_{\mathrm{em}}\left(\mu_{s}\right)} 
  \frac{Q_{q}\left(2 Q_{\ell}+Q_{q}\right)}{\beta_{0, \mathrm{em}}^{2}}
  \left(g_{0}\left(\eta_{\mathrm{em}}\right)+\frac{\alpha_{\mathrm{em}}
  \left(\mu_{s}\right)}{2 \pi} \beta_{0, \mathrm{em}} \ln \eta_{\mathrm{em}}
  \ln \frac{\omega}{\mu_{s}}\right)\right]
  \end{align}
where $g_{0}(x)\, =\, 1\, -\, x\, +\, \ln x$ and 
$\eta_{\mathrm{em}}\left(\mu_{s}, \mu\right) \, \equiv \,
\alpha_{\mathrm{em}}\left(\mu_{s}\right) / \alpha_{\mathrm{em}}(\mu)$.

In the LL approximation, the evolution function involving only final-state leptons is uniform from 
hard scale ($m_b$) to soft-collinear scale ($\mathrm{\Lambda_{QCD}}$) \cite{Beneke:2019slt},
  \begin{align}
  U_{\ell}\left(\mu_{b}, \mu\right) &\, =\, \exp\,  
  \left[\, \int_{\mu_{b}}^{\mu} \, \frac{d \mu^{\prime}}{\mu^{\prime}}\, 
  \Gamma_{\mathrm{c}}\left(\mu^{\prime}\right)\,  
  \ln \, \frac{m_{B_{q}}}{\mu^{\prime}}\, \right]\, ,
  \end{align}
and the anomalous dimensions
  \begin{align}
  \Gamma_{c}&\, =\, \frac{\alpha_{\mathrm{em}}}{\pi} \, 2\,  Q_{\ell}^{2}\, .
  \end{align}


\subsubsection{The resummed result}
\label{subsubsection4.2.3}

We collect at this point all evolution factors, including the evolution 
of hard function Eq.(\ref{Hardevolu}) and soft one Eq.(\ref{Softevolu}), 
to turn the factorized amplitude Eq.(\ref{Factorized amp}) to a 
resummed one, 
  \begin{align}\label{resummed resultO91}
  i\, \mathcal{A}_{9} \, =\, & T_{+}\left(\mu_{hc}\right)\,  m_{B_{q}} 
  \int_{0}^{1} d u \int_{0}^{\infty} d \omega \, U_{h}  (\mu_{b}, \mu_{hc} )\, U_{\ell}\left(\mu_{hc}, \mu_{s c}\right)\,
  U_{s}^{\mathrm{QED}} \left(\mu_{h c}, \mu_{s} ; \omega\right)  \, \nonumber\\
 &
  2\, H_{9}\left(u ; \mu_{b}\right)\, J_{m}\left(u ; \omega ; \mu_{h c}\right) 
  F_{B_{q}}\left(\mu_{h c}\right)  \phi_{+}\left(\omega ; \mu_{h c}\right) \, .
  \end{align}
Plugging the explicit result of $J_m$ in 
Eq.(\ref{TimeorderSCETIresult3}) and $T_{+}$ in 
Eq.(\ref{Tplus}) into Eq.(\ref{resummed resultO91}), 
we get 
  \begin{align}\label{Amplitudeloop}
  i \mathcal{A}_{9}\, =\, & 
  - \frac{i}{2} \frac{\alpha_{\mathrm{em}}\left(\mu_{hc}\right)}{4 \pi}
  Q_{\ell}Q_{q} m_{\ell} m_{B_{q}} F_{B_{q}} \mathcal{N}
  \left[\bar{\ell}_{sc} (1 + \gamma_5 ) \ell_{\overline{sc}} \right]\, 
  \int_{0}^{1} d u \bar u  \int_{0}^{\infty} \frac{d \omega}{\omega} \,
 U_{h}  (\mu_{b}, \mu_{hc} )\,   \nonumber\\
 &  
U_{\ell}\left(\mu_{hc}, \mu_{s c}\right)\,
 U_{s}^{\mathrm{QED}}\left(\mu_{h c}, \mu_{s} ; \omega\right) \,
 2\, H_{9} (u, \mu_{b}) \, \phi_{+}\left(\omega ; \mu_{h c}\right)\,
 \ln \left(1 + \frac{u}{\bar u} \frac{n_+ p_{\ell^-} \omega }{m_{\ell}^{2}} \right)\, .
 \end{align}
The term proportional to $C^{\mathrm{eff}}_7$ in hard function $H_{9} (u, \mu_{b})$ 
will not cause endpoint-singularity as $u \to 0$ when convoluted by jet function 
(eq.(\ref{TimeorderSCETIresult3})) in eq.(\ref{Amplitudeloop}) , 
which is different from the case appeared in $B_{d, \, s} \to \mu^+\, \mu^-$. 
In the effective theory, the virtualities of both the soft and collinear modes 
are suppressed, and the endpoint singularity might arises when we drop the power 
suppressed terms.  This is just the case for  $B \to \mu^+ \mu^-$ decay, 
where the final state muon(or anti-muon)  is collinear particle,  
thus the endpoint singularity in the $C_7^{\rm eff}$ term as $u \to 0$ 
arises from the hard-collinear and collinear convolution integral for the box diagrams.
However,  things are different in $B \to \tau^+ \tau^-$ decay,  where the final state $\tau$ 
is a hard-collinear particle, its virtuality is different from the soft quark inside the $B$ meson 
and cannot be neglected in the effective theory expansion.  
As a result, the $\tau$ mass appears in the jet function and there is no endpoint divergence 
when it is convoluted with the $1/u$ in the hard function. 
Therefore, the QED correction to $B \to \tau^+ \tau^-$ is in nature free from endpoint singularity,
 and this conclusion does not depend on the expansion by the QED coupling constant. 

The tree level contribution in $\alpha_{\mathrm{em}}$ from $Q_{10}$ 
in LL approximation is
  \begin{align}\label{Amplitudetree}
   i \, \mathcal{A}_{10}
  &\, =\, -\, i\, m_{\ell}\,  F_{B_{q}}\,  \mathcal{N} \, C_{10}\left(\mu_{b}\right)\,  
  U_{\ell}\left(\mu_{b}, \mu_{s c}\right)\, 
  \left[\, \bar{\ell}_{sc}\, \gamma_5\,  \ell_{\overline{sc}}\, \right]\, .
  \end{align}
With the addition of  the tree level contribution Eq.(\ref{Amplitudetree}), 
we write the total amplitude formally as
  \begin{align}
  i \, \mathcal{A}\, =\, -\, i\, 
  \left(\, A_{10}\, [\, \bar{\ell}_{sc}\, \gamma_5\,  \ell_{\overline{sc}}\, ]\, +\, 
A_{9}\,  [\, \bar{\ell}_{sc}\,(1\, +\,  \gamma_5)\,  \ell_{\overline{sc}}\, ]\, \right)\, ,
\end{align}
where the scalar reduced amplitudes $A_{9,\, 10}$ can be 
read from Eqs. (\ref{Amplitudeloop}) and (\ref{Amplitudetree}).

\section{Decay width with the addition of ultrasoft photons}
\label{ultrasoft}

Actually, the virtual QED correction to $B_q \, \to\,  \tau^+\,  \tau^-$ 
discussed up to now is not infrared safe. It is necessary to include 
real radiation as $\Gamma\left[\, B_{q}\, \rightarrow\, \ell^{+} \, \ell^{-}\, 
\right]\, +\, \Gamma\left[\, B_{q} \, \rightarrow\,  \ell^{+}\,  \ell^{-}\, +\, 
n \, \gamma\, \right]$ to guarantee the decay rate IR-finite and 
well-defined, where $n$ denotes the number of real radiation photon. 
The energy of real radiation $E_\gamma$ will be subject to the 
experimental setup in the form of a photon-energy cutoff $\Delta E$, 
$E_\gamma\,  <\, \Delta E$. Throughout we will restrict the discussion 
to the case of $\Delta E \, \ll\,  \Lambda_{\mathrm{QCD}}$ and contain 
an arbitrary number of additional ultrasoft real photons. It is possible to 
obtain the real ultrasoft contribution by matching 
$\mathrm{HQET}\times \mathrm{bHLET} $ at a soft scale $\mu_s$
to an effective theory that contains the electrically neutral $B$-meson 
field and heavy lepton fields with fixed velocity label $v_{\ell, \bar \ell}$, 
in analogy with heavy-quark effective theory. The ultrasoft fields can be 
decoupled from the heavy lepton fields, $\ell_{sc} \, \rightarrow \, 
S_{v_{\ell}}\,  \ell_{sc}^{(0)}$, where $S_{v_{\ell}}$ is ultrasoft Wilson 
line, and are not coupled to the neutral initial state at leading power 
in $1/m_b$. Therefore the amplitude can be factorized into the 
non-radiative amplitude $\mathcal A_i$ and an ultrasoft matrix 
element as follows,
  \begin{align}\label{usoft}
  \mathcal{N}_{\Delta B=1}\,  C_{i}\, 
  \left\langle\, \ell^+\,  \ell^- \, X_{s}\, \left|\, Q_{i}\, \right| \, \bar{B}_{q}\,
  \right\rangle
  \, =\, \mathcal{A}_{i}\, 
  \left\langle \, X_{s}\, \left|\, S_{v_{\ell}}^{\dagger}(0)\, S_{v_{\bar{\ell}}}(0)\, \right| 0\, 
  \right\rangle, 
  \quad i=9, 10\, ,
  \end{align}
where $X_{s}$ is an arbitrary ultrasoft state consisting of photons, 
and possibly electrons and positrons. $\mathcal A_i$ is the 
non-radiative amplitude as we discussed before. Formally, 
the matching of $\mathrm{HQET}$ 
with quark fields to ultrasoft EFT of point-like hadrons at the 
scale $\sim \Lambda_{\mathrm{QCD}}$ must be done 
non-perturbatively. However as the $B$-meson is neutral 
and decoupled in the far infrared, the case that the ultrasoft 
photon decoupled from final leptons in the ultrasoft EFT is 
similar to the SCET treatment of soft radiation from top-quark 
jets \cite{Fleming:2007qr,vonManteuffel:2014mva}.
The resummation of large logarithmic corrections of the
ultrasoft function with the RG technique can also be achieved 
by the ultrasoft EFT, in analogy with SCET treatment in
\cite{Fleming:2007qr}.

The partial decay width is obtained by squaring the full 
amplitude Eq.(\ref{usoft}) and summing over all ultrasoft 
final states with total energy less than $\Delta E$,
  \begin{align}\label{Twidth}
  \Gamma\left[\, B_{q} \, \rightarrow \, \tau^{+} \, \tau^{-}\, \right](\Delta E)
  \,  =\,  \frac{m_{B_{q}}}{8\,  \pi} \, \beta_{\tau}\, 
  \left(\, \left|A_{10}+A_{9} \right|^{2}\, +\, \beta_{\tau}^{2}\, |A_{9}|^{2}\, \right) \, 
  \mathcal{S}\left(v_{\ell}, v_{\bar{\ell}}, \Delta E\right) \, ,
  \end{align}
with $\beta_{\tau}\, =\, \sqrt{1\, -4\,  m_{\tau}^{2} / m_{B_{q}}^{2}}$.
The ultrasoft contribution $\mathcal{S}\left(v_{\ell}, v_{\bar{\ell}}, \Delta E\right)$ 
is
  \begin{align}
  \mathcal{S}\left(v_{\ell}, v_{\bar{\ell}}, \Delta E\right)
  \, =\, \sum_{X_{s}}\, \left|\, \left\langle \, X_{s}\, \left|\, S_{v_{\ell}}^{\dagger}(0) \, 
  S_{v_{\bar{\ell}}}(0)\, \right|\,  0\, \right\rangle\right|^{2} \, 
  \theta\left(\Delta E\, -\, E_{X_{s}}\right)\, ,
  \end{align}
with the one-loop soft function for massive final particles given 
in \cite{vonManteuffel:2014mva},
  \begin{align}
  S^{(1)}\left(v_{\ell}, \, v_{\bar{\ell}}, \, \Delta E\right)
  \, =\, \ln \, \frac{2\,  \Delta E}{\mu}\,  \gamma_{0}^{s}(x)\, +\, c_{1}(x)\, ,
  \end{align}
where 
  \begin{align}
  \gamma_{0}^{s}(x)&\, =\, -\, 8\, 
  \left[\, 1\, +\, \frac{1\, +\, x^{2}}{1\, -\, x^{2}} \, G(0 ; x)\, \right] \, ,\\
  c_{1}(x) \, =&\, 
  \left[\, \frac{1\, +\, x^{2}}{1\, -\, x^{2}}\left(-2\, G^{2}(0 ; x)\, +\, 
  8\, G(1 ; x)\, G(0 ; x)\, -\, 8 \,G(0,1 ; x)\, -\,  \frac{4 \pi^{2}}{3}\right)
  \right.\nonumber\\
  &\left.-4\, \frac{1\, +\, x}{1\, -\, x}\, G(0 ; x)\, \right] \, .
  \end{align}
Harmonic polylogarithms with n weights are defined as 
(see also Ref.~\cite{vonManteuffel:2014mva}),
  \begin{align}
  G\left(w_{1}, \ldots, w_{n} ; x\right)
  &\, =\, \int_{0}^{x} \, \frac{d t}{t-w_{1}}\,  G\left(w_{2}, \ldots, w_{n} ; t\right)\, ,
  \nonumber\\
  G(0, \ldots, 0 ; x)&\, =\, \frac{1}{n !} \, \ln ^{n}(x)\, ,
  \end{align}
for at least one of $\left\{w_{1}, \ldots, w_{n}\right\}$ different 
from zero and all $w_{i}\, =\, 0$, respectively, where
  \begin{align}
  x\, =\, \frac{1\, -\, \sqrt{1\, -\, \frac{4\,  m_{\tau}^{2}}{m_{B_{q}}^{2}}}}
  {1\, +\, \sqrt{1\, -\, \frac{4\,  m_{\tau}^{2}}{m_{B q}^{2}}}}\, .
  \end{align}
The resummed soft function can be achieved by using 
the QED exponentiation theorem as a approximate, 
e.g. full soft function can be considered as the exponent 
of the one-loop result, 
  \begin{align}\label{RadiativeFactor}
  \mathcal{S}\left(v_{\ell}, \, v_{\bar{\ell}}, \, \Delta E\right)
  \, =\, \exp\,  \left[\, \frac{\alpha_{\mathrm{em}}}{4\,  \pi} \, Q_{\ell}^{2}\,  
  S^{(1)}\left(v_{\ell}, \, v_{\bar{\ell}}, \, \Delta E\right)\, \right]\, .
  \end{align}

\section{Branching fractions of $B_q \, \to\,  \tau^+\,  \tau^-$}
\label{numerical}

We are now ready to numerically evaluate the non-radiative 
branching fractions for $B_q\,  \to\,  \tau^+\,  \tau^-$ defined 
as
  \begin{align}
  \mathrm{Br}_{q \tau}^{(0)} \, \equiv\,   \Gamma^{(0)}
  \left[\, B_{q} \, \rightarrow\,  \tau^{+}\,  \tau^{-}\, \right]\, \tau_{B_q}\, ,
  \end{align}
where the non-radiative width 
$ \Gamma^{(0)}\left[\, B_{q}\,  \rightarrow\,  \tau^{+} \, \tau^{-}\, \right]\,$
is just the width in Eq.(\ref{Twidth}) with the ultrosoft function 
$\mathcal{S}\left(v_{\ell},\,  v_{\bar{\ell}}, \, \Delta E\right)\, =\, 1$.
$ \tau_{B_q}\, $ is the lifetime of $B_q$ meson.
Our inputs are collected in Table \ref{tab:num-input}. 
  \begin{table}
  \centering
  \renewcommand{\arraystretch}{1.7}
  \resizebox{\columnwidth}{!}{
  \begin{tabular}{|lll|lll|}
  \hline
  Parameter
  & Value
  & Ref.
  &  Parameter
  & Value
  & Ref.
  \\
  \hline 
  $G_F$                            & $1.166379 \cdot 10^{-5}$ GeV$^{-2}$  &~\cite{Workman:2022ynf}
  & $m_Z$                            & $91.1876(21)$ GeV                    &~\cite{Workman:2022ynf}
  \\
  $\alpha_s^{(5)}(m_Z)$                & $0.1181(11)$                         &~\cite{Workman:2022ynf}
  & $m_\tau$                          & $1.77686(12)$ MeV                  &~\cite{Workman:2022ynf}
  \\
  $\alpha_{\mathrm{em}}^{(5)}(m_Z)$          &  $1/127.955(10)$ 
  &~\cite{Workman:2022ynf}
  & $m_t$                  &  $162.5^{+2.1}_{-1.5}$ GeV         &~\cite{Workman:2022ynf}
  \\
  \hline
  $ m_b(m_b)$             & $4.180(8)$ GeV                      &~\cite{Aoki:2019cca}
  & $ m_c(m_c)$                & $1.275(9)$ GeV                       &~\cite{Aoki:2019cca}
  \\
  $m_b^{\text{pole}}$  & $ 4.816(9)$ GeV           &
  &   $m_c^{\text{pole}}$  & $1.841(10)$ GeV           &
  \\
  \hline
  $\mu_{h}$  & $4.18$ GeV  &
  &$\mu_{hc}$  & $1.5$ GeV  &
  \\
  \hline
  $m_{B_s}$                        & $5366.92(10)$ MeV                    &~\cite{Workman:2022ynf}
  & $m_{B_d}$                        & $5279.66(12)$ MeV                    &~\cite{Workman:2022ynf}
  \\
  $f_{B_s}$                        & $230.3(1.3)$ MeV                     &~\cite{Aoki:2019cca} 
  & $f_{B_d}$                        & $190.0(1.3)$ MeV                     &~\cite{Aoki:2019cca}
  \\
  $\tau_{B_s}$                     & $1.520(5)$ ps                       &~\cite{Workman:2022ynf}
  & $\tau_{B_d}$                     & $1.519(4)$ ps                        &~\cite{Workman:2022ynf}
  \\
  $\lambda_{B_s}(\mu_0)$                & $400(150)$ MeV                       &
  & $\lambda_{B_d}(\mu_0)$                & $350(150)$ MeV                       &
  \\
  $\hat {\sigma}^{(1)}_{B_s}(\mu_0)$           & $0.0(0.7)$                           &~\cite{Beneke:2018wjp}
  & $\hat {\sigma}^{(1)}_{B_d}(\mu_0)$           & $0.0(0.7)$                           &~\cite{Beneke:2018wjp}
  \\
  $\hat {\sigma}^{(2)}_{B_s}(\mu_0)$           & $0(6)$                           &~\cite{Beneke:2018wjp}
  & $\hat {\sigma}^{(2)}_{B_d}(\mu_0)$           & $0(6)$                           &~\cite{Beneke:2018wjp}
  \\
  \hline
  $\lambda$                        & $0.22500(67)$                   &~\cite{Workman:2022ynf}
  & $\bar{\rho}$                      & $0.159(10)$       &~\cite{Workman:2022ynf}
  \\ 
  $A$                              & $0.826^{+0.018}_{-0.015}$       &~\cite{Workman:2022ynf}
  & $\bar{\eta}$                      & $0.348(10)$                    &~\cite{Workman:2022ynf}
  \\
  \hline
  \end{tabular}
  }
  \renewcommand{\arraystretch}{1}
  \caption{\label{tab:num-input}
  \small
   Numerical values for parameters: 
   $\alpha_{s} $ and $\alpha_{\mathrm{em}}$ are the 
   $\overline {\mathrm{MS}}$-renormalized coupling constants. 
   The masses of quarks, $m_t$, $m_b$ and $m_c$, are in 
  $\overline {\mathrm{MS}}$ scheme. $m_{b,c}^{\text{pole}}$ 
  are pole masses of bottom-quark and charm-quark, which are 
  obtained from the corresponding $\overline {\mathrm{MS}}$-values 
  and will be used in our numerical calculations. The values of the 
  Wilson coefficients at $\mu_b\, =\, 4.18$ GeV are 
  $C_{1-6}\, =\, \{-0.331,\, 1.010, \, -0.005,\, -0.090,\, 0.00038,\, 0.001\}$,  
   $C_{7}^{\mathrm{eff}}\, =\, -0.316$, $C_{9}\, =\, 4.200$ and 
   $C_{10}\, =\, -4.543$. The $B_{d, s}$ meson decay constants 
   $f_{B_{d, s}}$ are averages from the FLAG group for 
   $N_f \, =\,  2 + 1 + 1$ from 
   \cite{Bazavov:2017lyh, ETM:2016nbo, Dowdall:2013tga, Hughes:2017spc}. 
   The nonperturbative parameters entering the three-parameter model 
   for leading twist $B$-meson LCDA contain $ \lambda_{B_{d, s}}$, 
   $\hat {\sigma}^{(1)}_{B_{d, s}}$ and $\hat {\sigma}^{(2)}_{B_{d, s}}$.
  We will use Wolfenstein parametrization of the CKM matrix with the 
  Wolfenstein parameters, $\lambda$, $\bar{\rho}$, $A$ and $\bar{\eta}$.}
  \end{table}

We will apply the following three-parameter model for leading 
twist $B$-meson LCDA as used in \cite{Beneke:2018wjp,Shen:2020hsp,Shen:2021yhe,Lu:2022fgz,Qin:2022rlk},
  \begin{eqnarray}
  \phi_+(\omega)\, =\, {\Gamma(\beta)\over \Gamma(\alpha)}\, 
  {\omega\over \omega^2_0}\, \mathrm{e}^{-{\omega\over \omega_0}}\, 
  U\left(\beta-\alpha,\, 3-\alpha,\, {\omega\over \omega_0}\right)\, ,
  \end{eqnarray}
where $U(\alpha,\, \gamma,\, x)$ is the confluent hypergeometric 
function of the second kind. In the amplitude formula 
Eq.(\ref{Amplitudeloop}), only the first inverse moment 
and the logarithmic moments are needed, which are 
defined by \cite{Beneke:2011nf} 
(see also, for instance 
\cite{Bell:2013tfa,Feldmann:2014ika,Wang:2015vgv,Wang:2016qii,Wang:2018wfj,Shen:2020hfq,Liu:2020ydl,Wang:2021yrr,Galda:2022dhp,Beneke:2022msp,Cui:2022zwm})
  \begin{eqnarray}
  \label{moment}
  {1\over\lambda_B(\mu)}&\, =\, &
  \int_0^\infty\, {d\omega\over \omega}\, \phi_+(\omega)\, ,\\
  \sigma_n(\mu)&\, =\, &
  \lambda_B(\mu)\, \int_0^\infty\, {d\omega\over \omega}\, 
  \ln^n{\mu_0\over \omega}\, \phi_+(\omega) \, ,
  \end{eqnarray}
respectively. The first inverse moment $\lambda_B$ and 
the first two logarithmic moments $\hat \sigma_1$ and 
$\hat \sigma_2$ can be expressed as functions of 
three parameter $\alpha$, $\beta$ and $\omega_0$,  
  \begin{eqnarray}
  \lambda_B&\, =&\, 
  {\alpha\, -\, 1\over \beta\, -\, 1}\, \omega_0\, ,\\
  \hat \sigma_1&\, =&\, 
  \psi(\beta\, -\, 1)\, -\, \psi(\alpha\, -\, 1)\, +\, 
  \ln\, {\alpha\, -\, 1\over \beta\, -\, 1}\, , \\
  \hat \sigma_2&\, =&\, 
  {\pi^2\over 6}\, +\, \hat \sigma_1^2\, -\,
   [\, \psi'(\beta\, -\, 1)\, -\, \psi'(\alpha\, -\, 1)\, ]\, ,
  \end{eqnarray}
whose values are listed in the Table \ref{tab:num-input}. 
Apart from the SM and hadronic parameters listed in 
Table \ref{tab:num-input}, our results are also depend 
on two renormalization scales, hard scale $\mu_h$ 
used in the calculation of the Wilson coefficient and 
hard collinear scale $\mu_{hc}$ in the second step 
of matching. We choose their central values as 
$\mu_h\, =\, m_b\, =\, 4.18\, \mathrm{GeV}$ and  
$\mu_{hc}\, =\, 1.5\, \mathrm{GeV}$, which will be 
varied as $\mu_h\,  \in \, \{\, m_b/2,\, 2\, m_b\, \}$ 
and $\mu_{hc}\,  \in \, \{\, 1.5-0.5,\, 1.5+0.5\, \}$ 
to estimate errors of branching ratios.

The non-radiative branching fraction of 
$B_q\,  \to\,  \tau^+\, \tau^-$ for the central values 
of the parameters in Table \ref{tab:num-input} are
  \begin{align}
  \label{numericalBrd}
  \mathrm{Br}^{(0)}(B_d \to \tau^+ \tau^-)&\, =\, 
  \left(\, 1.993_{\text{\textcolor{gray}{(LO)}}} \, -\, 
  0.001_{ \text{\textcolor{gray}{(NLO)}}}\right)\, 
  \times\,  10^{-8}\, ,\\
  \label{numericalBrs}
  \mathrm{Br}^{(0)}(B_s \to \tau^+ \tau^-)
  &\, =\, \left(\, 6.940_{\text{\textcolor{gray}{(LO)}}}  \, -\, 
  0.003_{\text{\textcolor{gray}{(NLO)}}}\right)\, \times \, 10^{-7}\, ,
  \end{align}
where the first and second terms in r.h.s. of 
Eqs.(\ref{numericalBrd}) and (\ref{numericalBrs}) are results 
from the leading order of $\alpha_{\mathrm{em}}$, and from 
the QED and QCD corrections to the next to leading order 
and leading logarithmic accuracy, respectively. The numerical 
value of the QED and QCD corrections lead to an overall 
enhancement of the branching fraction of approximately $0.04\%$, 
which is much smaller than the one in $B_q\,  \to\,  \mu^+\, \mu^-$ 
case (overall reduction about $0.5\%$). The reason is that, even though 
the power enhancement effect $m_b/ \Lambda_{\mathrm{QCD}}$ also 
appear in $\tau$ case, the single-logarithm term from hard-collinear 
function in Eq.(\ref{TimeorderSCETIresult3}) for $\tau$ final states 
are not as large as in $\mu$ case (the single-logarithmic enhancement 
of order $\ln\, m_{b}\, \Lambda_{\mathrm{QCD}} / m_{\mu}^{2} \, \sim\, 5$ 
for the $C_{9}^{\text {eff }}$ term \cite{Beneke:2017vpq}) due to 
hard-collinear scale mass of $\tau$.

For completeness, we consider uncertainties of the non-radiative
 branching fractions in Eqs.(\ref{numericalerrorBd}) and 
 (\ref{numericalerrorBs}). They arise from $B_q$ meson decay 
constants $f_{B_q}$ and $B_q$ meson LCDA parameters, 
$\lambda_{B_{q}}$, $\hat {\sigma}^{(1)}_{B_{q}}$ and 
$\hat {\sigma}^{(2)}_{B_{q}}$, the SM parameters including
$m_t$ and $m_b^{\mathrm{pole}}$, and two renormalization 
scales, $\mu_h$ and $\mu_{hc}$, which have been added in 
quadrature in the following as
  \begin{align}
  \label{numericalerrorBd}
  \mathrm{Br}^{(0)}(B_d \to \tau^+ \tau^-)&\, =\, 
  \left(\, 1.992\, ^{+0.174}_{-0.258}\, \right) \, \times \, 10^{-8}\, ,\\
  \label{numericalerrorBs}
  \mathrm{Br}^{(0)}(B_s \to \tau^+ \tau^-)&\, =\, 
  \left(\, 6.937\, ^{+1.454}_{-2.313}\, \right) \, \times\,  10^{-7}\, .
  \end{align}

Finally, the branching fraction for the infrared-finite observables 
of $B_{q} \, \rightarrow\, \ell^{+} \, \ell^{-}\, (n\, \gamma)$ with 
ultrasoft photon energy $E_\gamma\,  < \, \Delta E$ will be 
given by multiplied with the soft-photon exponentiation factor 
in Eq.(\ref{RadiativeFactor}). $E_\gamma$ would depend on 
how well the real photon could be detected by a particular 
experiment. Current experiment analyses, such as LHCb, 
use dilepton energy cuts that would correspond to an allowable 
soft photon of up to 60 MeV \cite{LHCb:2013vgu,LHCb:2012skj}.  
We can also choose the same signal window, 
$\Delta E\,  \simeq\,  60$ MeV, for 
$B_{q}\,  \rightarrow\,  \tau^{+} \, \tau^{-}\, (n\, \gamma)$ 
and the numerical results are
  \begin{align}\label{numericaltotal}
  \mathrm{Br} (B_d \, \to \, \tau^+\,  \tau^-) (\Delta E)
  &\, =\left(\, 1.960 \,^{+0.171}_{-0.253} \, \right)\times\,  10^{-8}\, ,\\
  \mathrm{Br} (B_s \, \to \, \tau^+\,  \tau^-)(\Delta E)
  &\, =\left(\, 6.822\, ^{+1.425}_{-2.267}\, \right) \times \, 10^{-7}\, ,
  \end{align}
which mean that the radiative factor in Eq.(\ref{RadiativeFactor}) 
for $B_{q}\,  \rightarrow \, \tau^{+}\,  \tau^{-}$ is about $98\%$ of 
the non-radiative rate.

\section{Summary}
\label{Summary}

We have considered the QED corrections to $B_{q}\, \to\, \tau^{+} \, \tau^{-}$ 
at next-to-leading order in $\alpha_{\mathrm{em}}$ and leading logarithmic 
resummation under the framework of SCET. The ultrasoft real photons are 
treated in the limit of static heavy leptons and decoupled from heavy leptonic 
fields, which means the ultrasoft QED effect can be factorized from 
nonradiative correction, and is universal for $B_{q}\,  \to \, \ell^{+}\,  \ell^{-}$ 
with $\ell=e,\, \mu,\, \tau$. 
The treatment for this effect on $B_{q}\,  \to\,  \tau^{+} \, \tau^{-}$ is same as 
on $B_{q}\,  \to\,  \mu^{+} \, \mu^{-}$, and similar to the SCET treatment of 
soft radiation in top-quark jets. Then we concentrate on virtual QED effects 
which are from the process-specific energy scales set by the external 
kinematics and internal dynamics of $B_{q} \, \to\,  \tau^{+} \, \tau^{-}$. 
We have performed two steps of matching from $\mathrm{QCD} \times 
\mathrm{QED}$ onto $\mathrm{SCET_{I}}$ and subsequently onto 
$\mathrm{HQET} \times \mathrm{bHLET} $. Hard fluctuations from 
$m_b$ scale are integrated out in the matching onto $\mathrm{SCET_{I}}$ 
and successively (anti-)hard-collinear fluctuations with 
$m_b \, \Lambda_{\mathrm{QCD}}$ virtualities are decoupled from 
$\mathrm{HQET} \times \mathrm{bHLET} $. Different from muon 
leptonic $B$ decays, the effective operator in $\mathrm{SCET_{I}}$ 
for $B_{q}\,  \to\,  \tau^{+} \, \tau^{-}$ is only $\mathcal O_9$ as the 
$Q_7$ can be matched to $\mathcal O_9$ by integrating out hard 
photon from the electromagnetic dipole operator. For completeness 
of QED corrections, we calculate the hard functions at NLO although 
they are not relevant to power enhanced effects. 
In $\mathrm{HQET} \times \mathrm{bHLET} $, there is only so-called 
$A1$-type operator for $B_{q}\, \to\, \tau^{+}\, \tau^{-}$. By matching the 
time-ordered product of the operator $\mathcal O_9$ together with two 
Lagrangians $\mathcal{L}_{\xi q}^{(1)}$ and $\mathcal{L}_{m}^{(0)}$ 
to the matrix element of  $A1$-type operator in $\mathrm{HQET} \times \mathrm{bHLET} $, 
we integrate the (anti-)hard-collinear virtualities of 
spectator-quark, which lead to formally power enhanced effects by a 
factor $m_b/\Lambda_{\mathrm{QCD}}$ as discussed in $B_{q} \, \to \, 
\mu^{+}\,  \mu^{-}$. However, for $B_{q} \, \to \, \tau^+\,  \tau^-$, as the 
mass of tau is just the order of hard-collinear scale, the logarithm term 
arising from the contribution of hard-collinear photon and lepton virtuality 
in the second matching step is small, which would not induce large 
enhanced QED effects as in muon case even though the same power 
enhancement term appears in $B_{q}\,  \to\,  \tau^+\,  \tau^-$. Numerically, 
together with the resummation at the leading logarithm accuracy in the 
both QCD and QED coupling, the values of the QED and QCD corrections 
lead to an overall enhancement of the branching fraction of approximately 
$0.04\%$, compared with overall reduction of branching fraction about 
$0.5\%$ in $B_q \, \to\,  \mu^+\, \mu^-$ case. 

\subsection*{Acknowledgements}
We are grateful to Yu-Ming Wang for a resultful cooperation and very 
valuable discussions. S.-H. Zhou also thanks Yan-Bing Wei for useful discussions. 
The research of Y. L. Shen is supported by the National 
Natural Science Foundation of China with Grant No.12175218 and the 
Natural Science Foundation of Shandong with Grant No. ZR2020MA093. 
S. H. Zhou acknowledges support from the National Natural Science 
Foundation of China with Grants No.12105148.


\bibliographystyle{JHEP} 
\bibliography{refs.bib}
\end{document}